\documentclass[nolinenumbers, twocolumn]{aastex631}

\newcommand{\chas}{CH$\alpha$S}

\usepackage{amsmath}

\shorttitle{\chas}
\shortauthors{Melso et al.}

\graphicspath{{./}{figures/}}

\begin{document}

\title{The Circumgalactic H$\alpha$ Spectrograph (\chas) \\ I. Design, Engineering, and Early Commissioning}

\correspondingauthor{Nicole Melso}
\email{nmelso@arizona.edu}

\author[0000-0002-4895-6592]{Nicole Melso}
\affiliation{Department of Astronomy, Columbia University, 550 W. 120th Street MC 5246, New York, NY 10027, USA}
\affiliation{Columbia Astrophysics Laboratory, Columbia University, 550 W. 120th St. MC 5247, New York, NY 10027, USA} 
\affiliation{University of Arizona, Steward Observatory, 933 N. Cherry Avenue, Tucson, AZ 85721, USA}

\author{David Schiminovich}
\affiliation{Department of Astronomy, Columbia University, 550 W. 120th Street MC 5246, New York, NY 10027, USA}
\affiliation{Columbia Astrophysics Laboratory, Columbia University, 550 W. 120th St. MC 5247, New York, NY 10027, USA}

\author{Brian Smiley}
\affiliation{Columbia Astrophysics Laboratory, Columbia University, 550 W. 120th St. MC 5247, New York, NY 10027, USA}

\author{Hwei Ru Ong}
\affiliation{Columbia Astrophysics Laboratory, Columbia University, 550 W. 120th St. MC 5247, New York, NY 10027, USA}

\author[0000-0002-1321-3748]{B\'{a}rbara Cruvinel Santiago}
\affiliation{Department of Physics, Columbia University, 538 W. 120th Street 704 Pupin Hall MC 5255, New York, NY 10027, USA}
\affiliation{Columbia Astrophysics Laboratory, Columbia University, 550 W. 120th St. MC 5247, New York, NY 10027, USA}

\author[0000-0001-7714-6137]{Meghna Sitaram}
\affiliation{Department of Astronomy, Columbia University, 550 W. 120th Street MC 5246, New York, NY 10027, USA}
\affiliation{Columbia Astrophysics Laboratory, Columbia University, 550 W. 120th St. MC 5247, New York, NY 10027, USA}

\author{Ignacio Cevallos Aleman}
\affiliation{Department of Physics, Columbia University, 538 W. 120th Street 704 Pupin Hall MC 5255, New York, NY 10027, USA}
\affiliation{Columbia Astrophysics Laboratory, Columbia University, 550 W. 120th St. MC 5247, New York, NY 10027, USA}

\author[0000-0003-3553-4144]{Sarah Graber}
\affiliation{Department of Astronomy, Columbia University, 550 W. 120th Street MC 5246, New York, NY 10027, USA}
\affiliation{Columbia Astrophysics Laboratory, Columbia University, 550 W. 120th St. MC 5247, New York, NY 10027, USA}

\author[0000-0002-3572-1049]{Marisa Murillo}
\affiliation{Department of Engineering, Columbia University, 500 W. 120th Street $\#$510, New York, NY 10027}
\affiliation{Columbia Astrophysics Laboratory, Columbia University, 550 W. 120th St. MC 5247, New York, NY 10027, USA}

\author{Marni Rosenthal} 
\affiliation{Department of Physics $\&$ Astronomy, Barnard College, 3009 Broadway, Altschul Hall 504A, New York, NY 10027}
\affiliation{Columbia Astrophysics Laboratory, Columbia University, 550 W. 120th St. MC 5247, New York, NY 10027, USA}

\author{Ioana Stelea}
\affiliation{Department of Astronomy, Columbia University, 550 W. 120th Street MC 5246, New York, NY 10027, USA}
\affiliation{Columbia Astrophysics Laboratory, Columbia University, 550 W. 120th St. MC 5247, New York, NY 10027, USA}

\begin{abstract}

The Circumgalactic H$\alpha$ Spectrograph (CH$\alpha$S) is a ground-based optical integral field spectrograph designed to detect ultra-faint extended emission from diffuse ionized gas in the nearby universe. CH$\alpha$S is particularly well suited for making a direct detection of tenuous H$\alpha$ emission from the circumgalactic medium (CGM) surrounding low-redshift galaxies. It efficiently maps large regions of the CGM in a single exposure, targeting nearby galaxies (d $< 35 $ Mpc) where the CGM is expected to fill the field of view. We are commissioning CH$\alpha$S as a facility instrument at MDM Observatory. CH$\alpha$S is deployed in the focal plane of the Hiltner 2.4-meter telescope, utilizing nearly all of the telescope's unvignetted focal plane (10 arcmin) to conduct wide-field spectroscopic imaging. The catadioptric design provides excellent wide-field imaging performance.  CH$\alpha$S is a pupil-imaging  spectrograph employing a microlens array to divide the field of view into $> 60,000$ spectra. CH$\alpha$S achieves an angular resolution of $[1.3 - 2.8]$ arcseconds and a resolving power of R$ = [10,000 - 20,000]$. Accordingly, the spectrograph can resolve structure on the scale of $1-5$ kpc (at 10 Mpc) and measure velocities down to 15-30 km/s. CH$\alpha$S intentionally operates over a narrow (30 $\AA$) bandpass; however, it is configured to adjust the central wavelength and target a broad range of optical emission lines individually. A high diffraction efficiency VPH grating ensures high throughput across configurations. CH$\alpha$S maintains a high grasp and moderate spectral resolution, providing an ideal combination for mapping  discrete, ultra-low  surface brightness emission on the order of a few milli-Raleigh. 
\end{abstract}


\section{Introduction} \label{sec:intro}

Where are the atoms in the Universe? This question highlights the last remaining frontier in our study and census of the evolving distribution of baryonic matter. The answer remains elusive due to the challenges of detecting gas at its lowest densities. A true mapping of diffuse emission requires observations that can reach exceedingly low surface brightnesses. Pioneering work with Fabry-Perot, deep long-slit, and integral-field unit (IFU) spectroscopy (e.g. \citealp{Vogel1995}) has sought to measure faint light from intergalactic clouds and galaxy halos. This remains very much a field in its early stages, requiring novel instrumentation that pushes the boundaries of ultra-low surface brightness spectroscopy. 

To meet this challenge, we have designed the Circumgalactic H$\alpha$ Spectrograph (CH$\alpha$S), an IFU spectrograph tailored to measure faint emission from diffuse gas in the local universe.  The primary science goal for CH$\alpha$S is to observe the circumgalactic medium (CGM) around low-redshift galaxies. As an integral field spectrograph, CH$\alpha$S can simultaneously map the mass, distribution, and kinematics of ionized gas undergoing cooling or recombination. Observations with CH$\alpha$S will address fundamental questions, investigating how much baryonic mass surrounds nearby spiral galaxies, how this matter is distributed, and how gas flows between the circumgalactic medium and the galactic disk. 

CH$\alpha$S occupies an important new parameter space in the design of low-surface brightness integral field spectrographs; it is a fast, ultra-narrowband, moderate-resolution spectrograph with exceptional grasp. CH$\alpha$S collects over 60,000 simultaneous spectra with a band-width optimized to study the relevant dynamics of low-redshift galaxies and their surrounding gaseous halos. Integral field spectroscopy is well-suited for faint, diffuse observations because it offers unconstrained pointing, wide-field spatial sampling, and superior survey speed \citep{Bershady2009, Bacon2017, Morrissey2018}. CH$\alpha$S is able to compete in survey speed with IFU spectrographs on 8-10 meter class telescopes at a fraction of the cost.  

Many of the CH$\alpha$S design goals are motivated by the observational and theoretical properties of ionized gas in the low-redshift CGM. These properties drive the following scientific requirements.

\begin{enumerate}
    \item Target a broad range of optical emission lines: H$\alpha$ ($\lambda 6563$), H$\beta$ ($\lambda  4861$), [S II] ($\lambda 6718$), [N II] ($\lambda 6584$), [O III] ($\lambda 5007$), [O II] ($\lambda 3727$), [O I] ($\lambda 6300$)
    \item Cover the extended physical size of the CGM ($50 - 100$ kpc) in the local universe (d $< 35$ Mpc) while resolving individual structures/features on scales of 0.1-1 kpc at a distance of 10 Mpc.
    \item Coverage over a continuous systemic velocity range from $\sim (0 - 2400]$ km/s (z $< 0.008$) with the ability to measure rest frame Doppler kinematics as large as $\sim \pm 600$ km/s down to a velocity resolution of 15-30 km/s. 
    \item Achieve sensitivity to emission on the order of $\simeq$ 1-10 mR (approximately $0.1 \% - 1\%$ the sky background intensity). 
\end{enumerate}

Several sources and ionization mechanisms contribute to the ionization of the CGM \citep{Shull2012, Haardt2012, Bland-Hawthorn2017}. Emission radiation traces the recombination of photoionized gas and the recombination/cooling of collisionally ionized gas and collisionally excited gas. The namesake emission signature for CH$\alpha$S is the prominent Balmer H$\alpha$ line; however, CH$\alpha$S targets a broad range of optical emission lines. Spectral line ratios can be used to map the efficacy of different ionization mechanisms throughout the galactic halo \citep{Baldwin1981}. 

The CGM is a massive gas reservoir extending hundreds of kpc from the disk of a galaxy out to the viral radius and beyond \citep{Shull2014,  Borthakur2016}. In order to cover the extended physical size of the CGM in a single exposure, CH$\alpha$S is optimized to operate over a wide field of view ($10' \times 10'$) set by the observatory telescope$-$detector pairing. The spatial resolution within the CH$\alpha$S FOV is motivated by the expected gas distribution and the size of the small-scale substructure that populates the halo. Observations and simulations of the CGM are often in agreement that cicumgalactic gas distribution is patchy \citep{Borthakur2016, Oppenheimer2018, Lehner2020} and hosts many sources of multi-scale (0.01 - 1000 kpc) structure including small clumps/clouds \citep{Churchill2003, Stocke2013, Crighton2015}, radial gradients \citep{Borthakur2016}, and large scale filaments \citep{Haffner1998, Churchill2012, Joung2012}. CH$\alpha$S is designed to capture radial/spatial dependencies in the extended gas distribution while resolving features on the average length-scale of substructure in the CGM.

The CH$\alpha$S spectral window is tuned to detect circumgalactic gas kinematically associated with low-redshift galaxies ($z < 0.008$) up to a systemic velocity of 2400 km/s. Despite the variety of dynamic components in the halos of these galaxies  \citep{Martin2012, Rubin2012, Ho2014, Ho2017,Zheng2017, RodrigezdelPino2019, Zabl2019, French2020, Ho2020}, gas within the virial radius is typically gravitationaly bound with CGM velocities not exceeding the escape velocity (550 km/s for Milky Way mass galaxies) \citep{Tumlinson2017}.  Accordingly, CH$\alpha$S aims to measure Doppler kinematics as large as $\pm 600 \rm \ km \ s^{-1}$ in a given filter ($\sim 25 \ \AA$), encompassing the majority of gas in the low-redshift CGM of Milky-Way-like galaxies. The velocity resolution selected for CH$\alpha$S ($15-30 \rm \ km \ s^{-1}$) is on par with typical spectral line widths observed in the CGM \citep{Werk2013}, which are limited by thermal broadening and non-thermal turbulent motions on the order of $\lesssim  20-50 \rm \ km \ s^{-1}$ (\cite{Tumlinson2017} and references therein).

Very few direct detections of CGM emission have been made in the low-redshift regime beyond the Milky Way halo \citep{Bland-Hawthorn1997, Fumagalli2017}. One of the main obstacles impeding a direct detection of warm/warm-hot ionized gas in the CGM is the diffuse nature of the gas ($10^{-6} < n_{H} < 10^{-2}$) \citep{Tumlinson2017}, producing an extremely tenuous emission signal on the order of $\sim 1-10$ mR \citep{Corlies2016, Zhang2016}. This signal is easily overpowered by sky background or detector noise. CH$\alpha$S is designed to detect emission with a sensitivity of to $0.1 \% - 1\%$ of the sky-background intensity \footnote{Dependent on the filter bandpass ($\rm I_{sky} = 2 \ R \AA^{-1})$} \citep{Leinert1998}. 

CH$\alpha$S is a dedicated instrument for the Hiltner 2.4-meter telescope at MDM Observatory on Kitt Peak, bringing full-field spectral imaging capabilities to the observatory. The MDM facilities are owned and operated by a consortium consisting of Dartmouth College, Columbia University, Ohio State University, Ohio University, and the University of Michigan. CH$\alpha$S will be a permanent facility instrument at MDM Observatory, available to the consortium. This is the debut paper for the Circumgalactic H$\alpha$ Spectrograph. We present the optical design in Section \ref{sec:opticaldesign}, the mechanical design in Section \ref{sec:mechdesign}, the expected performance in Section \ref{sec:performance}, and the early commissioning results in Section \ref{sec:earlycomm}. We conclude with a discussion of the spectrograph optimization in Section \ref{sec:discussion} and a summary of key points from this publication in Section \ref{sec:summary}.

\section{Optomechanical Design}
\label{sec:opticaldesign}

\begin{figure*}[ht]
\begin{center}
\includegraphics[width=0.8\textwidth]{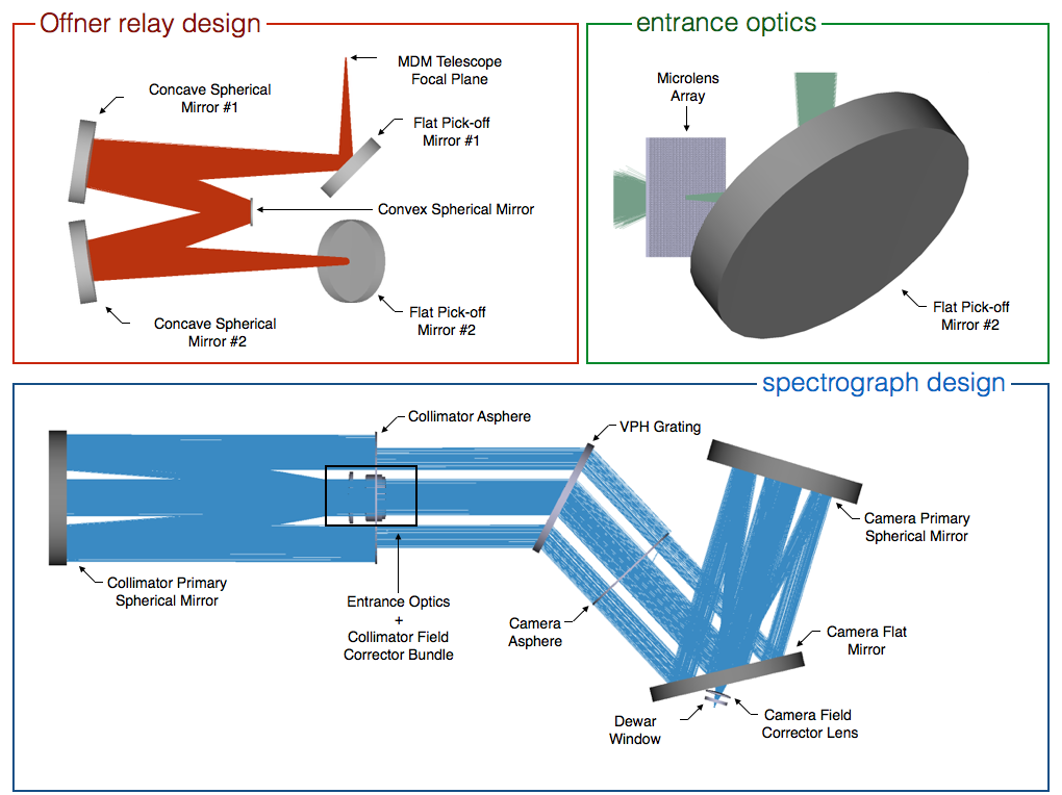}
\caption{Zemax ray-trace of the CH$\alpha$S design. In this figure, we split the design into three subsets: the relay optics, the entrance optics, and the spectrograph optics. The total sequential optical path through all three is: (relay optics) telescope focal plane, flat pick-off $\# 1$, concave spherical $\# 1$, convex spherical, concave spherical $\# 2$, (entrance optics) flat pick-off $\# 2$, microlens array,  (spectrograph optics) collimator field corrector bundle, collimator primary, collimator asphere, grating, camera asphere, camera flat, camera primary, camera field corrector, detector.  \label{fig:summary}}
\end{center}
\end{figure*}

The CH$\alpha$S optomechanical design is summarized in Figure \ref{fig:summary} and Figure \ref{fig:optmechsummary}. While the CH$\alpha$S performance and diffraction efficiency were originally optimized for H$\alpha$ emission, reduced chromaticity in the catadioptric design and the narrow instrument bandpass have allowed us extend the operational wavelength to nearly the full optical range. The CH$\alpha$S design triumphs in its configuration flexibility and in the cost-savings afforded by the incorporation of many commercially available optics.

\subsection{Focal Plane Optics} 
\label{optics:focal}

CH$\alpha$S sits in the focal plane of the MDM Hiltner 2.4-m telescope. It attaches to the telescope Multi-Instrument System (MIS), utilizing the finder module (with calibration lamps) and guider module. CH$\alpha$S has a separate shutter installed near the focal plane of the telescope \footnote{Uniblitz CS65 shutter and D880C driver}, which is triggered by the detector exposure commands through the MDM 4K detector electronics box. A system of relay optics is used to extend and re-image the telescope focal plane down into the entrance of the spectrograph with enough clearance to avoid collision between the MIS and the CH$\alpha$S collimator. The Offner relay design used in CH$\alpha$S can be seen in Figure \ref{fig:summary}. It consists of two identical concave spherical mirrors, one convex spherical mirror, and two flat pick-off mirrors, which in total relay the focal plane by $\sim$1.5 m with a nearly one-to-one mapping ($\rm m = -1.001265$). All of the optics are coated with a UV Enhanced Protective Silver High Reflective Coating ($>97 \% \ @ \ 400-1000$ nm) applied by Teledyne Acton Optics. The final flat in the relay is positioned to redirect the re-imaged focal plane into the spectrograph entrance optics.

\subsection{Microlens Array}
\label{sec:microlens}
The entrance to the spectrograph is a microlens array, which sits directly in the re-imaged focal plane exiting the relay system. The lenslets themselves re-image the pupil of the telescope and act as a focal reducer, converting the F/7.5 telescope input to an F/2.5 output that can be fed into the fast collimator design. We have purchased two custom mirolens arrays from Advanced Microoptics Systems (a$\mu$s): one array with 250 $\mu$m pitch lenslets and one array with 125 $\mu$m pitch lenslets. Hereafter, these are referred to as the ML250 array and ML125 array. The ML250 and ML125 arrays have $> 60,000$ and $> 240,0000$ individual lenslets respectively, creating a powerful integral field spectrograph with high spatial and spectral resolution.

We present the CH$\alpha$S microlens array parameters in Table \ref{table:microlens}. Each array is a $68 \times 68$ mm clear aperture Fused Silica optic composed of hexagonal plano-convex lenslets, covering nearly the full field of view of the Hiltner 2.4-m telescope. An anti-reflection (AR) coating is applied to the planar side of each array with a reflectance of $<1\%$ between 400 nm and 900 nm. On the convex side of the array, a black chromium mask is applied between the lenslets, filling in the gaps where the lenslets intersect and creating a circular aperture around each hexagonal lenslet profile. The fused silica and black chromium mask on this side has a $< 4\%-5\%$ reflectance between 400 nm and 900 nm.  This mask is used to stop down the lenslet aperture slightly and to keep light from scattering at the lenselet interstices. An additional field mask can be installed next to the microlens array to block a specific pattern of lenslets, masking bright regions and maximizing the spectral filling factor at the detector. 

As seen in the non-sequential ray trace in Figure \ref{fig:microlens}, each plano-convex array can be flipped 180 degrees in the optical path without altering the imaging capabilities. In both configurations the focal plane of the telescope is incident on the side of the array with optical power. Both microlens arrays were designed with F/2.5 lenslets. They are slightly slower than the F/2.2 collimator in order to reduce the amount of light lost off the edges of the grating. For plano-convex lenslets, the effective focal length is independent of the array thickness, and each array is 2 mm thick to avoid deformation. When calculating the effective focal length, each lenslet can be approximated as a thick lens re-imaging the telescope pupil via refraction. We incorporate aberration effects and calculate the nominal best focus position of the lenslet array via non-sequential ray trace. This positioning can be fine-tuned using the motorized collimator focus and telescope focus.

CH$\alpha$S is a pupil imaging spectrograph. It uses a microlens array to sample the focal plane, as was originally done in the design of the TIGER IFS \citep{Bacon2001, Courtes1982}. The CH$\alpha$S microlens array covers the full field of view, and each lenslet bins the focused field positions from a small patch of sky. The exit pupil of each lenslet is a demagnified image of the telescope pupil \citep{Courtes1982}, referred to as a micropupil. The microlens array creates a grid of micropupils that form the focal plane of the collimator and the entrance to the spectrograph (Figure \ref{fig:microlens}). Since the telescope pupil (not the field) is re-imaged at each lenslet, imaging information within each spatial resolution element is not preserved and the PSF is independent of spatial field variation \citep{Bacon1995}. Without the diffraction grating, these micropupils are directly re-imaged onto the detector with demagnification through the collimator/camera optics (Figure \ref{fig:microlens}).  With the grating in place, the space between micropupils can be filled with the dispersed spectra. A camera aligned with the dispersion direction images the spectra on the detector. Overlap between adjacent spectra restricts the spectral bandpass, and a narrowband filter is used to shorten the spectra. Slightly rotating the lenslets with respect to the dispersion direction allows us to lengthens the vertical distance to the next micropupil/spectrum \citep{Bacon2001}. With a slight rotation of approximately 5-6 degrees, each spectrum now grazes past its immediate neighbors, improving spectral packing and extending the allowable bandpass.

\begin{figure*}[ht]
\label{fig:microlens}
\begin{center}
\includegraphics[width=0.85\textwidth]{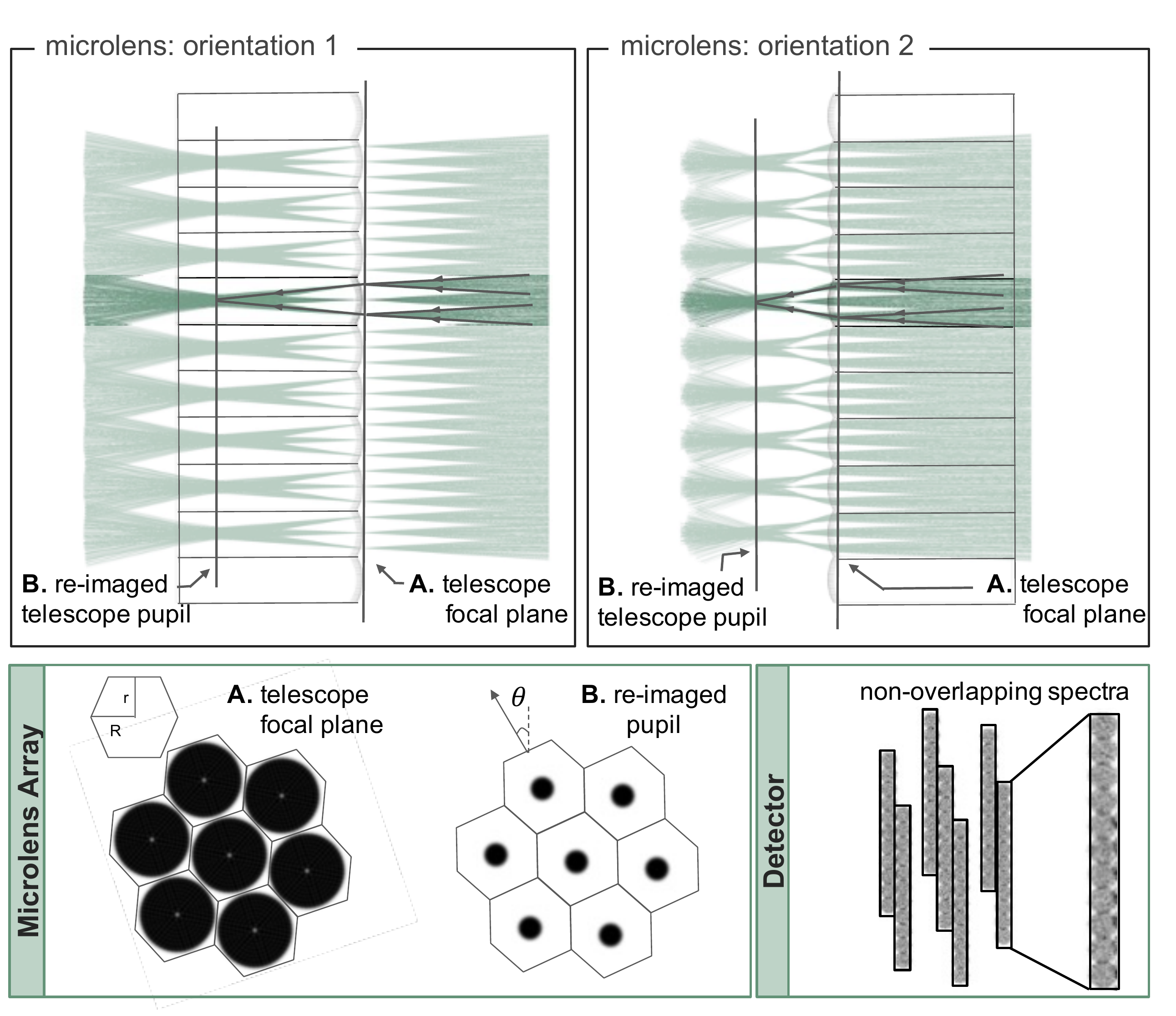}
\caption{The top panels show non-sequential Zemax ray-trace showing a section of the CH$\alpha$S plano-convex microlens array in two orientations: (left) convex side facing towards the telescope focal plane and (right) convex side facing away from the telescope focal plane. The array can be flipped 180 degrees in the optical path without altering the imaging capability.  Orientation 1 re-images the telescope pupil inside the array. Orientation 2 re-images the telescope pupil outside the array, allowing us to access it and add additional masking. The bottom panels show non-sequential ray trace imaging at the microlens array and at the detector. The bottom left panel shows the geometric pupil of the telescope in the telescope focal plane. The plano-convex lenslets re-image the telescope pupil, resulting in the micro-pupils shown in the bottom middle panel. This re-imaging/demagnification adds space between lenslets for the spectra to disperse. At the detector (bottom right), the spectra extend vertically and do not overlap for many resolution elements. \label{fig:microlens}}
\end{center}
\end{figure*}

\begin{deluxetable}{ccccccBcccccc}
\tablecaption{Microlens array parameters as manufactured by Advanced Microoptics Systems (A$\mu$S). Effective focal lengths are calculated using the thick lens approximation at a central wavelength of $\lambda = 0.6583$ nm for low-redshift H$\alpha$. Both arrays are made of Fused Silica (n = 1.4563 @ 0.6583 nm).}
\label{table:microlens}
\tablehead{
\colhead{Pitch} & \colhead{Aperture} & \colhead{Radius} &
\colhead{Eff Focal} & \colhead{F/$\#$} & 
\colhead{Thickness}\\
\colhead{(mm)} & \colhead{(mm)} & \colhead{(mm$^{-1}$)} & \colhead{(mm)} & 
\colhead{} & \colhead{(mm)}  
}
\startdata
0.25 & 0.225 $\pm$ 0.001 & 0.257$\pm 3\%$ & 0.563 & 2.5 & 2.0\\
0.125 & 0.113$\pm$ 0.001 & 0.129 $\pm 3\%$ & 0.283 & 2.5 & 2.0 \\
\enddata
\end{deluxetable}

\begin{deluxetable}{cccccBcccccBccccc}
\tablecaption{Optical performance of spectrograph convolved with lenslet pupil imaging.}
\label{table:optical}
\tablehead{
\colhead{Pitch} & \colhead{RMS Radius} & \colhead{GEO Radius} &  FWHM & sigma \\
\colhead{} & {Zemax } &  \colhead{Zemax}  & \colhead{Measured} & \colhead{Measured} \\
\colhead{(mm)} & {($\mu$m)} &  \colhead{($\mu$m)}  & \colhead{($\mu$m)} & \colhead{($\mu$m)}
}
\startdata
0.25 & 19.0 & 27.3 & 57 & 24\\
0.125 & 8.9 & 13.5 & 30.5 & 13 \\
\enddata
\end{deluxetable}

\subsection{Collimator $\&$ Camera}
\label{subsec:colcam}

CH$\alpha$S has a Schmidt collimator and a Schmidt camera. This design achieves excellent wide-field optical performance while utilizing many commercially available components and leveraging the comparably low-cost manufacturing of optical spheres. The diverging beam from from the microlens array is collimated by a Celestron 36 cm Rowe-Ackermann Schimdt Astrograph (RASA). Among commercially available telescopes, the RASA has a large aperture (356 mm), and a fast (f/2.2) wide-field ($4.3^{\circ}$ FOV) design that combined result in an impressively high \'etendue. The optical performance is a $< 6.3$ micron RMS spot size across the full FOV and spanning the full optical bandpass. 

The CH$\alpha$S camera is a custom folding Schmidt design consisting of an aspheric corrector plate, a folding flat mirror, and a spherical primary mirror. The asphere at the camera entrance is the mass-produced Celestron 28 cm RASA corrector plate \footnote{the same as the C12 Schmidt-Cassegrain corrector plate}. The folding flat is a custom optic manufactured by Sydor, and the primary is a custom sphere manufactured by Optimax. The folding Schmidt design places the detector behind the flat outside of the optical path, significantly reducing obscuration. The flat has a cut machined out of the center to accommodate the return beam and pass light through to the detector. The CH$\alpha$S camera design achieves an excellent optical performance, producing RMS spot sizes of $< 10$ microns over a $10 \times 10$ arcmin FOV and a wide range of central wavelengths spanning $[400 - 900]$ nm.

We note that the CH$\alpha$S camera design is shorter than a classical Schmidt telescope; the distance from the sphere to the corrector plate is less than the radius of curvature. While certain optimizations of the shortened Schmidt design can achieve abberation correction with a singlet self-achromatic\footnote{The chromatic focal shift at the front and back surfaces of the lens are equal and opposite} fused silica field corrector \citep{Wynne1977}, the CH$\alpha$S field corrector has a small focal shift on the order of a micron over the full optical bandpass [400- 900 nm]. The self-achromatic shortened Schmidt optimization is constrained by the distance between the asphere and the sphere, which for CH$\alpha$S is partially restricted by the folding geometry. Moving the flat closer to the primary places it further upstream in the return beam, requiring a larger cutout that introduces additional obscuration. In the CH$\alpha$S camera design we minimize this obscuration, as the axial achromatic shift within each narrowband (3 nm) observation is negligible. The obscuration from the flat is less than the entrance optics obscuration, which is approximately 30$\%$ of the stop aperture.  

\subsection{VPH Grating}
\label{sec:diffraction}
The CH$\alpha$S diffraction element is a volume phase holographic (VPH) grating mounted on a rotational stage between the collimator and camera. Many instruments swap between a collection of VPH gratings in order to optimize the diffraction efficiency in different wavelength regimes. In contrast, CH$\alpha$S operates over a wide range of central wavelengths $[400 - 900]$ nm with exceptional efficiency using only a single VPH grating. The peak diffraction efficiency is shifted to the desired central wavelength by simply rotating the grating to adjust the angle of incidence \citep{Barden2000, Baldry2004}. While having a wide field of view and a broad spectral range usually poses a challenge for VPH gratings, coupling CH$\alpha$S with a narrow bandpass filter allows us to operate away from the grating's central optimization with incredibly high throughput.

The CH$\alpha$S VPH grating was manufactured by Wasatch Photonics with a fringe frequency of $\nu$ = 1200 lines/mm. This contributes to the moderate resolving power (R $\sim 10,000 - 20,000$) and sets a collimator-camera angle that ranges from $\sim [35^{\circ} - 65^{\circ}]$ depending on selected wavelength. The grating is a large-format optic [340 mm x 290 mm x 20 mm] with an elliptical clear aperture. It is capped using fused silica substrates, each polished to excellent surface quality (mean $\rm PV = 1.197/RMS = 0.21$2 over a 98 mm aperture) and coated with a broadband ($400 - 1000$ nm) anti-reflective coating ($< 2\%$ reflectance for Bragg incidence at 658 nm). Grating parameters are summarized in Table \ref{table:gratingparams}.

VPH gratings diffract light using a thin layer of gelatin with the index of refraction modulated to form fringes. The diffraction efficiency is moderated by Bragg reflection; light is coherently diffracted with high efficiency when scattering in the gelatin volume creates constructive interference \citep{Baldry2004}. The Bragg condition \footnote{Parameters in the gelatin have subscript 2 ($\alpha_{2}$, $\beta_{2}$, $n_{2}$, $\Delta n_{2}$). Parameters satisfying the Bragg condition have a subscript b ($\alpha_{2b}$, $\lambda_{b}$).} (Equation \ref{eq:bragggrating}) is dependent on the fringe frequency in the grating ($\nu_{2}$), the wavelength of light ($\lambda$), and the angle of incidence in the grating gelatin with respect to the fringe structure ($\alpha_{2b}$). Therefore, the diffraction efficiency at a given wavelength can be maximized by selecting an angle of incidence that satisfies the Bragg condition.

\begin{equation}
    m\nu_{2}\lambda_{2b} = 2 n_{2}\sin \alpha_{2b}
\label{eq:bragggrating}
\end{equation}

\begin{deluxetable*}{ccccccccccccc}
\tablecaption{The grating and camera angles for key wavelengths in the three different channels (blue, red, and infrared). The Bragg angle is measured inside the gelatin relative to the grating fringes. The fringe slant is $\phi = 4.5^{\circ}$ The angle of incidence and angle of diffraction are measured in air relative to grating normal. The camera angle is the sum of the angle of incidence and angle of diffraction measured relative to the optical axis of the collimator. \label{table:gratingparams}}
\tablehead{
\multicolumn{1}{c}{Wavelength} & 
\multicolumn{1}{c}{Bragg} & 
\multicolumn{2}{c}{Incidence} & 
\multicolumn{2}{c}{Diffraction} & 
\multicolumn{1}{c}{Camera} &
\multicolumn{2}{c}{Anamorphic} &
\multicolumn{2}{c}{Angular Dispersion}&
\multicolumn{2}{c}{Linear Dispersion \tablenotemark{a}}\\ 
& & $+\phi$ & $-\phi$ & $+\phi$ & $-\phi$ &  & $+\phi$ & $-\phi$ & $+\phi$ & $-\phi$ & $+\phi$ & $-\phi$ 
}
\startdata
490 nm & 12.6$^{\circ}$ & 23.4$^{\circ}$ & 11.0$^{\circ}$ & -11.0$^{\circ}$ & -23.4$^{\circ}$ & 34.4$^{\circ}$ & 0.94$^{\circ}$ & 1.07$^{\circ}$ & $\rm 25'' \AA^{-1}$ & $\rm 27'' \AA^{-1}$ & $\rm 72 \ \mu m \ \AA^{-1}$ & $\rm 77 \ \mu m  \ \AA^{-1}$\\
658 nm & 17.0$^{\circ}$  & 29.7$^{\circ}$ & 17.0 $^{\circ}$ & -17.1$^{\circ}$ & -29.8$^{\circ}$ & 46.8$^{\circ}$ & 0.91$^{\circ}$ & 1.10$^{\circ}$ & $\rm 26'' \AA^{-1}$ & $\rm 28'' \AA^{-1}$ & $\rm 74 \ \mu m \ \AA^{-1}$ & $\rm 81 \ \mu m  \ \AA^{-1}$ \\
 658 nm \tablenotemark{b}  & 32.0$^{\circ}$ & 49.9$^{\circ}$ & 41.7$^{\circ}$ & -41.7$^{\circ}$ & -49.9$^{\circ}$ & 91.6$^{\circ}$ & 0.86$^{\circ}$ & 1.16$^{\circ}$ & $\rm 33'' \AA^{-1}$ & $\rm 38' \AA^{-1}$ & $\rm 95 \ \mu m \ \AA^{-1}$ & $\rm 110 \ \mu m  \ \AA^{-1}$ \\
900 nm & 23.6$^{\circ}$  & 39.5$^{\circ}$ & 26.2$^{\circ}$ & -26.4$^{\circ}$ & -39.7$^{\circ}$ & 65.9$^{\circ}$ & 0.86$^{\circ}$ & 1.17$^{\circ}$ & $\rm 28'' \AA^{-1}$ & $\rm 32'' \AA^{-1}$ & $\rm 79 \ \mu m \ \AA^{-1}$ & $\rm 92 \ \mu m  \ \AA^{-1}$ \\
\enddata 
\tablenotetext{a}{ 30 $\rm \mu m$ per pixel in 2x2 detector binning}
\tablenotetext{b}{ high dispersion mode using alternate VPH grating with a fringe frequency of 2173 lines $\rm mm^{-1}$ and a fringe slant of 2.5$^{\circ}$}
\end{deluxetable*}

\begin{figure}[ht]
\begin{center}
\includegraphics[width=1.06\columnwidth]{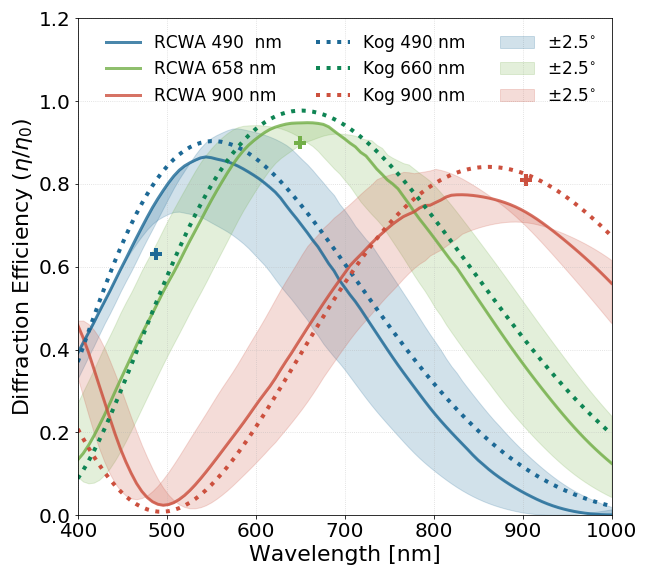}
\caption{ \label{fig:gratingeff} The CH$\alpha$S VPH grating diffraction efficiency (average polarization) as a function of wavelength at a fixed angle of incidence. The solid lines are rigorous coupled wave analysis (RCWA) carried out at Bragg incidence for central wavelengths of 490 nm (blue), 660 nm (green), and 900 nm (red). The same RCWA analysis is done for off-axis incidence angles with a Bragg deviation of $\pm 2.5$ degrees creating the angular Bragg envelope shown by the shaded regions. The dotted lines are the average lossless Kogelnik efficiency predictions again performed at Bragg incidence for 490 nm (blue), 660 nm (green) and 900 nm (red). The plus symbols are the manufacturer measured first order diffraction efficiencies at three distinct laser wavelengths (488 nm, 650 nm, 904 nm) taken using a 98 mm aperture and averaged for measurements across the grating. The RCWA analysis and measured data products in this plot were supplied by Wasatch Phonics.}
\end{center}
\end{figure}

Figure \ref{fig:gratingeff} shows the predicted (average polarization) CH$\alpha$S diffraction efficiency. The solid lines are rigorous coupled wave analysis (RCWA) carried out at Bragg incidence for central wavelengths of 490 nm (blue), 660 nm (green), and 900 nm (red). We note that over the narrow CH$\alpha$S bandpass ($\sim 3$ nm) there is very little change in the spectral efficiency response. On axis, CH$\alpha$S is always operating close to the Bragg wavelength, and the performance is not particularly dependent on the shape of the spectral efficiency response (spectral Bragg envelope) \cite{Kogelnik1969}.  Instead, CH$\alpha$S is more sensitive to deviations from the Bragg angle of incidence. Off-axis field positions covering the 10 arcmin CH$\alpha$S FOV undergo angular magnification through the fast, wide-field collimator optics. In the spectral direction, this angular magnification results in a $\pm 2.5$ degree spread in off-Bragg incidence angles encountered by the grating, which shifts the Bragg wavelength by $\pm 70$ nm at a central wavelength $\lambda = 660$ nm (well outside the $\sim 3$ nm filter bandpass) \citep{Robertson2000}. This can be seen in Figure \ref{fig:gratingeff} where RCWA analysis is done at the same central wavelengths for off-axis incidence angles with a Bragg deviation of $\pm 2.5$ degrees, creating the efficiency response shown by the shaded regions. The off-axis diffraction efficiency at the central wavelength decreases according to the profile of the angular Bragg envelope. Within this envelope the grating still maintains high throughput and the large off-Bragg acceptance angle allows us to use a fast collimator with high off-axis angular magnification over a wide field of view. 

VPH gratings have a number of additional tuning parameters inherent to the gelatin layer including the gelatin thickness (d), the average index of refraction in the gelatin ($\rm n_{2}$), and the index of refraction modulation amplitude ($\rm \Delta n_{2}$) that creates the fringe pattern.
In the CH$\alpha$S grating we select a thin gelatin depth ($\rm d_{eff} = 3\mu m$) and a high index of refraction modulation amplitude ($\Delta n = 0.11$) in order to maximize the angular bandpass ($\Delta \alpha \propto \frac {1}{\nu_{g} d}$) and spectral bandpass ($\frac{\Delta \lambda}{\lambda} \propto \frac{\cot \alpha_{B}}{\nu_{g} d}$) while still ensuring that the diffraction efficiency peaks around a central wavelength of 660 nm (Equation \ref{eq:vphdesign}) \citep{Barden2000}. 
 
 \begin{equation}
\label{eq:vphdesign}
   \Delta n_{2} d \approx \frac{\lambda}{2}
\end{equation}

The importance of the tuning parameters on diffraction efficiency is particularly interesting for an instrument like CH$\alpha$S, as we intend to use the grating far from its original optimization. A related feature in Figure \ref{fig:gratingeff} is the offset between the peak wavelength (sometimes called the blaze wavelength) and the Bragg wavelength. This occurs when we rotate the grating to operate away from the overall optimization at 660 nm. For example, the solid blue line represents RCWA efficiency predictions at Bragg incidence for a central wavelength of 490 nm, but the curve actually peaks at a blaze wavelength of 530 nm. This is a feature at Bragg incidence and is not an off-axis shift in the Bragg wavelength. To center the efficiency response on 490 nm at Bragg incidence, the ideal $\Delta n_{2}$ should be $\sim 0.08$ (Equation \ref{eq:vphdesign}). The fringe structure in the CH$\alpha$S grating is over-modulated (high $\Delta n_{2}$) for bluer wavelengths and under-modulated for redder wavelengths, manifesting as a shift in the efficiency response and an offset between the blaze peak and the Bragg wavelength (Wasatch private correspondence). If CH$\alpha$S had a wider bandpass and the spectral efficiency response was a more important factor in the overall diffraction efficiency, we could rotate the grating to a Bragg angle corresponding to a blaze wavelength peak at 490 nm. However, with the narrow instrument bandpass, we achieve a much smaller spread in the angular efficiency response by operating close to Bragg incidence and accepting an offset between the Bragg wavelength and blaze wavelength. Looking at the shaded off-Bragg angular efficiency response for 490 nm and 900 nm, the angular efficiency curves converge at the Bragg wavelength (where the deviation from off-Bragg incidence angle is zero). For fixed wavelength very close to the Bragg wavelength, the efficiency response as a function of field angle (in the spectral direction) is minimized and symmetric. Operating at the blaze wavelength would cause an asymmetric efficiency response at opposite field angles with a larger spread in the relative efficiency across the FOV. 

The final parameter in the CH$\alpha$S VPH grating design is the fringe slant, similar to the blaze angle in a ruled grating. The CHaS grating has a manufactured fringe slant of $\pm 4.5^{\circ}$. Depending on how the grating is installed in the optical path, the fringe slant can be either positive or negative, although the sign of the slant will affect the spectral resolution (see Table \ref{table:gratingparams} and discussion in Section \ref{sec:specperformance}).

In summary, the CH$\alpha$S VPH grating boasts a high diffraction efficiency of $>70\%$ for narrowband observations over the full field of view spanning central wavelengths from [490 - 900] nm. In addition, the grating performance is particularly superb at 660 nm near the H$\alpha$ line where we achieve a full-field diffraction efficiency of $> 90\%$. The plus markers in Figure \ref{fig:gratingeff} are the manufacturer measured first order diffraction efficiencies at three distinct laser wavelengths (488 nm, 650 nm, 904 nm) averaged across the grating aperture. These confirm the predicted on-axis efficiency close to the central wavelength.

\subsection{Mechanical Design}
\label{sec:mechdesign}

Part of the intent behind CH$\alpha$S, and the motivation for incorporating many commercial optics, is to develop a design that can be replicated or scaled for future applications. To make this feasible, we also require a mechanical design that is relatively modular, simple to construct, and cost effective. Here we detail the commercial and custom aspects of the CH$\alpha$S mechanical design. 

\begin{figure*}[ht]
\begin{center}
\includegraphics[width=0.8\textwidth]{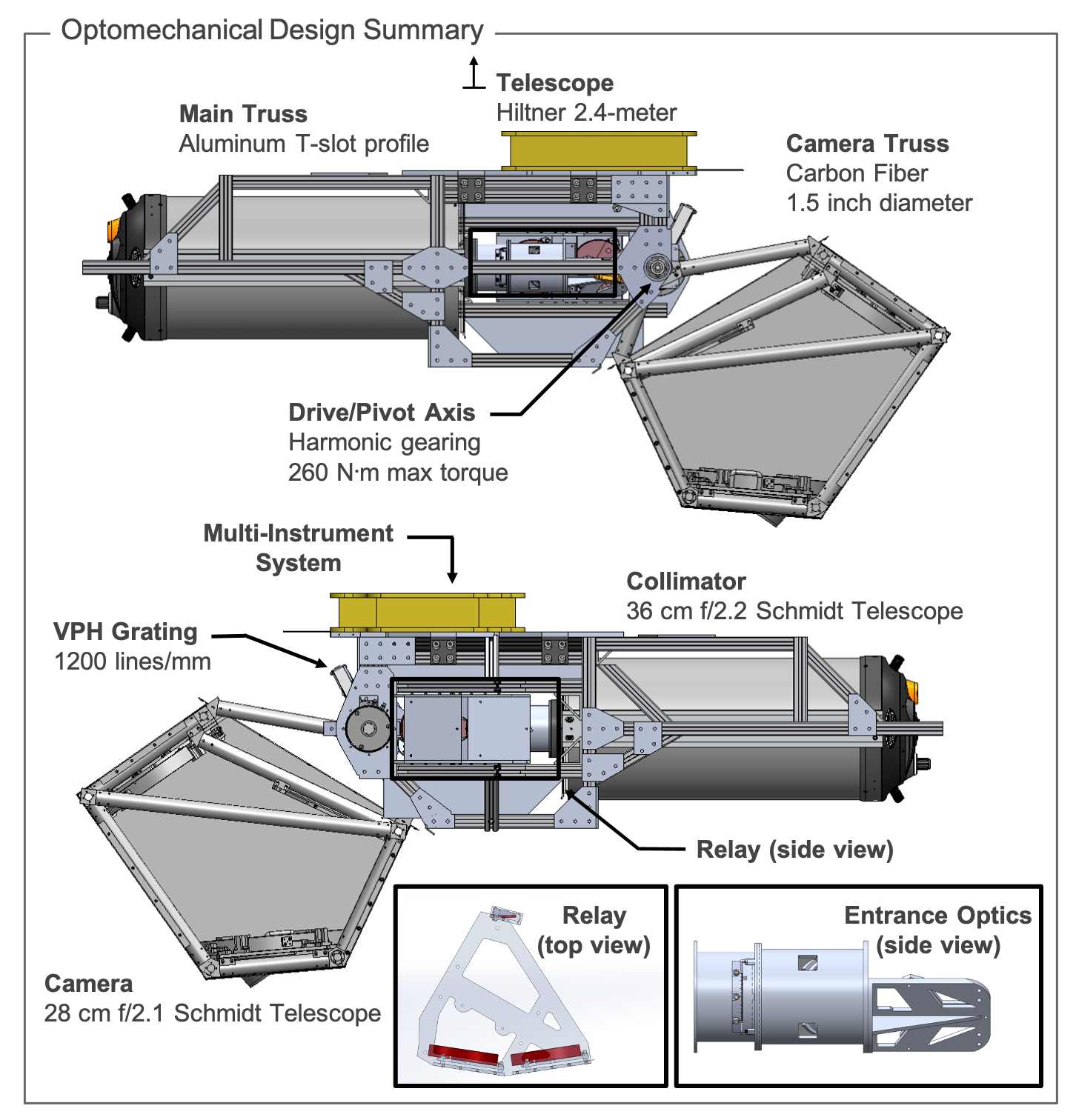}
\caption{The CH$\alpha$S optomechanical design modeled in SolidWorks. The top view is the East side of the spectrograph and the bottom view is the West side. CH$\alpha$S is primarily composed of two trusses. The main truss houses the relay, focal plane, and collimator optics. The camera truss houses the custom camera optics. The majority of the mechanical frame is assembled from prefabricated joint connector kits. The main truss is built from the 80-20 T-slot Aluminum Building System and the camera truss employs the Rock West Composites CARBONNect carbon fiber system. The camera and grating rotate about the labeled pivot axis. The relay optics are mounted on the side of the main truss.  \label{fig:optmechsummary}}
\end{center}
\end{figure*}

CH$\alpha$S is primarily composed of two trusses connected at a pivot point. The main truss houses the focal plane optics, relay system, collimator optics, and grating. It is designed to minimize flexure by keeping the center of gravity close to the telescope focal plane. The camera truss houses the custom camera optics. The majority of the mechanical frame is assembled from prefabricated joint connector kits. The main truss is built from the 80-20 T-slot Aluminum Building System and the camera truss employs the Rock West Composites CARBONNect carbon fiber system. Both the main truss and the camera truss are enclosed with black anodized sheet metal. A custom fabric bellows manufactured by Nabell connects the main truss enclosure to the camera truss enclosure, completing the light-tight envelope around the instrument optical path. 

The main truss and the camera truss are connected at a pivot point shown in Figure \ref{fig:optmechsummary}. This is the mounting location for the grating and the axis around which the camera and grating rotate independently. The camera drive rotation stage is a Harmonic Drive FHA25-US250-160-BL with a fail-safe brake and rotary position sensor. This motor was chosen to satisfy our mechanical requirements; namely, the ability to support the camera load with a 260 ft-lb max torque and zero backlash precision.  The grating angle is controlled separately from the camera angle using a Newport rotation stage URS100BPP. This motor meets our $20^{\prime\prime}$ bi-directional repeatability requirement, allowing us to position the grating angle to within one resolution element on the detector (two resolution elements with the ML125 array) (See Table \ref{table:gratingparams}). In order to minimize torque on the Camera Drive, the center of mass of the camera truss is kept as close as possible to the drive axis. The current operation employs counterweight arms, which are attached to the camera truss whenever the drive is engaged.  Once the camera is at the proper angle, these counterweight arms are detached replaced with rigid bracing arms that maintain position without the significant added mass of the counterweight.

\begin{figure*}[ht]
\begin{center}
\includegraphics[width=0.8\textwidth]{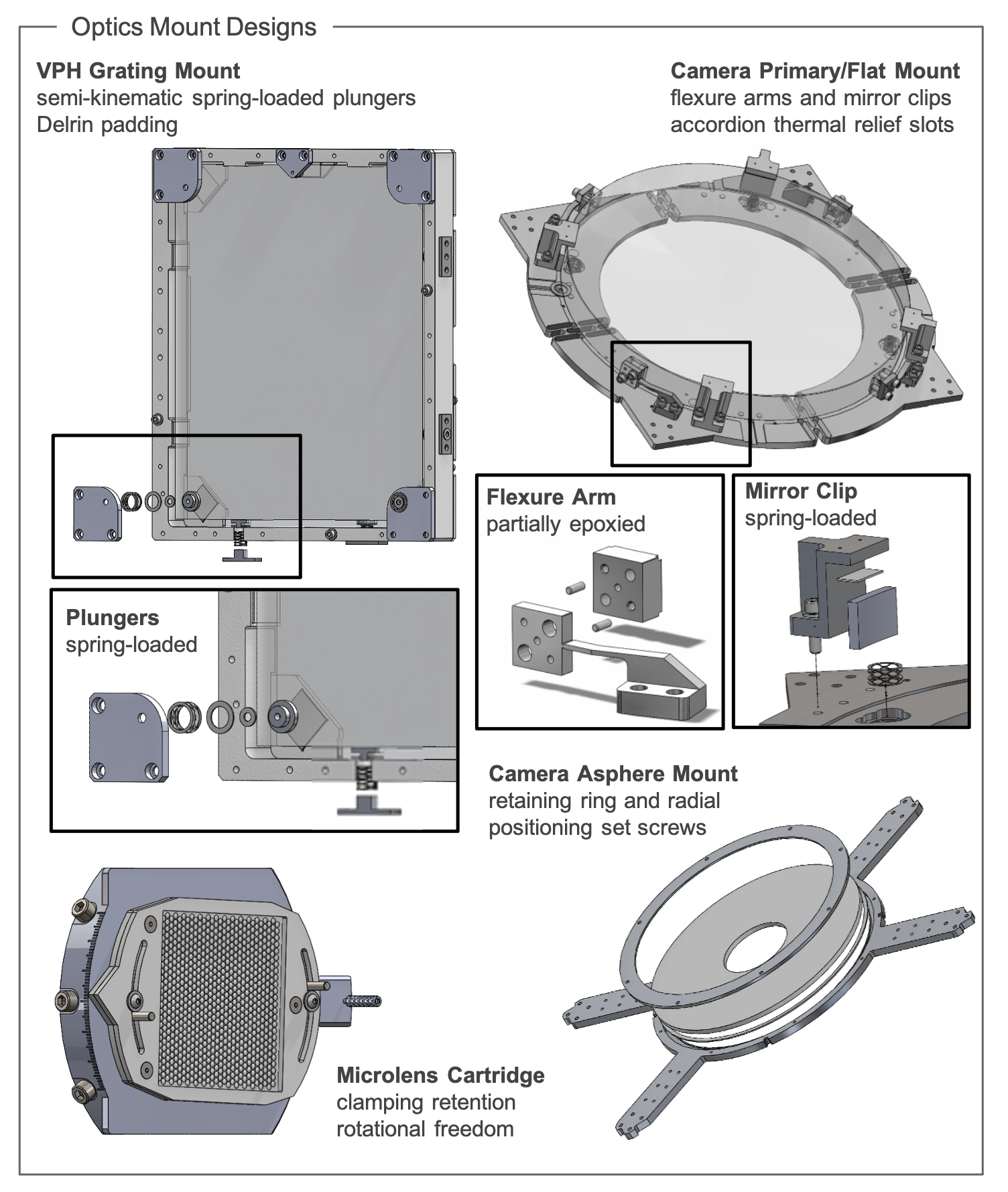}
\caption{We have designed custom mounts for the entrance optics, grating, and camera optics. Many of these optics are retained using spring-loaded or clamped mounting systems so that they can be removed from their mounts without the need to break any epoxy bonds. Here we show an overview of the custom mount designs in SolidWorks. Zoom in regions are exploded views of retention mechanisms.  \label{fig:mounts}}
\end{center}
\end{figure*}

We have designed custom mounts for the entrance optics, grating, and camera optics. An overview of the custom mount designs can be seen in Figure \ref{fig:mounts}.  Many of these optics are retained using a spring-loaded mounting systems in order to avoid transferring mount deformation to the optic or placing unwanted stress on the glass substrates. This design also allows us to remove the optics from their mounts without the need to break any epoxy bonds. Both the microlens array and grating are mounted such that they can be reversed in the optical path. In cases where the optics can be removed, reference points are used to ensure repeatable positioning. All of the optics mounts have built in tolerance for shimming. Due to the difference in the coefficient of thermal expansion (CTE) between the aluminum mounts and carbon fiber camera frame, both of the camera mirror mounts have accordion-like relief slots cut out to allow for differential compression and expansion during natural thermal cycling. The collimator optics are left installed in the Celestron optical tube, which is attached to our main truss using its built-in dovetail rails. All smaller mirrors in the relay/focal plane are bonded directly to small aluminum mounting plates with epoxy.

\section{Theoretical Performance}
\label{sec:performance}
We discuss the expected instrument performance assessed in the design stage via computer modeling and analytic calculations. These performance metrics map to the scientific requirements presented in Section \ref{sec:intro}. We have been working to confirm the expected performance through laboratory testing, calibration measurements, and engineering/commissioning time at the observatory (Section \ref{sec:earlycomm}). 

\subsection{Point Spread Function}
\label{subsec:opticalperformance}

The point spread function of CH$\alpha$S depends on the imaging properties of three subsystems: the telescope, the relay, and the spectrograph. The angular resolution of these subsystems are summarized in Table \ref{table:performance}. The spectrograph imaging performance further depends on the microlens array, collimator optics, diffraction grating, and camera optics. In this section, the theoretical optical performance and PSF of the spectrograph are calculated analytically as well as modeled with Zemax sequential and non-sequential ray tracing. 

We begin by reviewing the imaging properties of the spectrograph without the microlens array. It is important that the underlying PSF be significantly smaller than the object size at the entrance, or the re-imaged telescope pupil spots generated by the microlens array; this assures that the imaging performance is primarily set by the properties of the microlens array and not degraded by the spectrograph optics. We have designed CH$\alpha$S so that the spectrograph optics have an underlying PSF with a geometric radius of $\rm < 10 \ \mu m$ in all three channels across the full field of view. 

\begin{figure}[ht]
\begin{center}
\includegraphics[width=\columnwidth]{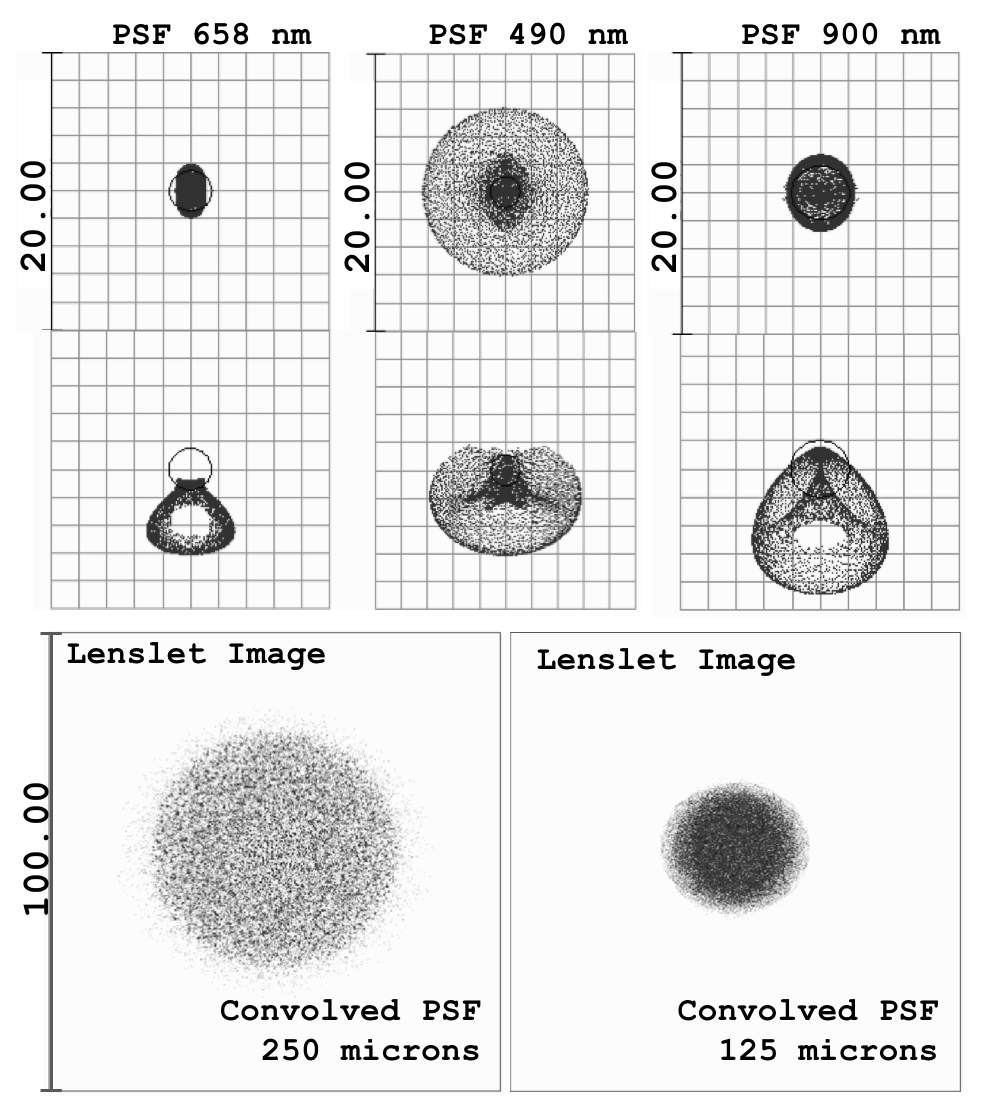}
\caption{ \label{fig:opticalperformance}Zemax ray trace showing the optical performance of the spectrograph (telescope and relay not included). The top panels are the unconvolved PSF (without the microlens array) in each of the three wavelength configurations. The bottom panels are the convolved pupil images (with the microlens array) for each lenslet pitch at a central wavelength of 658 nm. }
\end{center}
\end{figure}

With the microlens array inserted at the entrance to the spectrograph, each lenslet generates a pupil image. The pupil image is demagnified by the focal reduction in the lenslet,
\begin{equation}
\label{eq:focal_reduction}
w^{\prime} = \frac{F_{len}}{F_{tel}}w
\end{equation}
further demagnified by the camera-to-collimator focal ratio,
\begin{equation}
\label{eq:focal_reduction2}
w^{\prime \prime} = \frac{f_{cam}}{f_{col}}w^{\prime} = \frac{f_{cam}}{f_{col}}\frac{F_{len}}{F_{tel}}w
\end{equation}
and convolved with the spectrograph PSF (Figure \ref{fig:opticalperformance}) before being imaged on the detector. Using Equation \ref{eq:focal_reduction} and the clear aperture of the masked lenslets (Table \ref{table:microlens}), each pupil image has a radius of $37.5 \ \mu m$ with the ML250 array and $19 \ \mu m$ with the ML125 array. Assuming both the pupil image and the $10 \ \mu m$ spectrograph PSF are Gaussian, the blurred radius adds in quadrature and we expect geometric spot radii of approximately $30 \ \mu m$ and $17.5 \ \mu m$ for the ML250 and ML125 arrays respectively. Using Equation \ref{eq:focal_reduction2} to account for demagnification through the spectrograph, we calculate that each pupil image on the detector should have a geometric radius of approximately $28 \ \mu m$ for the ML250 array and $14 \ \mu m$ for the ML125 array. 

Zemax modeling of spectrograph PSF without the microlens array is shown by the (unconvolved) spot diagrams in the upper panel of Figure \ref{fig:opticalperformance}. We see that the Schmidt collimator and camera design achieves our desired wide-field imaging performance of $\rm < 10 \ \mu m$ in all three channels on and off axis. Modeling the lenslet and spectrograph together in Zemax at a wavelength of 658 nm, we generate the convolved pupil images shown in the bottom panel of Figure \ref{fig:opticalperformance}. 
 From these models we measure an expected geometric radius of 27 microns (19 microns RMS radius) for the ML250 array. The ML125 array model produces a 13.5 micron geometric radius (9 micron RMS radius). These values are on par with the analytic estimates above. We summarize these models in Table \ref{table:optical} alongside the optical performance measurements made during early commissioning (described in Section \ref{sec:earlycomm}). 

\subsection{Spectral Resolution}
\label{sec:specperformance}

CH$\alpha$S targets individual emission lines within the optical range of 400-900 nm (Section \ref{sec:intro}), intentionally limiting the instrument operational bandpass to 3 nm using a narrowband filter. Each lenslet is dispersed into a short spectrum with limited overlap. The dispersion at the detector is expected to be $\rm 0.4 \ \AA \ pix^{-1}$ at 658 nm (See Table \ref{table:gratingparams}), and each ML250 array spectral resolution element has a FWHM of about $45 \ \mu m$ or 1.5 pixels. Accordingly, CH$\alpha$S has a spectral resolution of approximately $\rm 0.6 \ \AA$ corresponding to a velocity resolution of 27 $\rm km \ s^{-1}$. The ML125 array with smaller pitch lenslets cuts the size of each spectral resolution element approximately in half, which allows us to distinguish emission features down to a velocity resolution of 14 $\rm km \ s^{-1}$. 

Many aspects of the CH$\alpha$S design factor into the spectroscopic performance of the instrument. As seen in Equation \ref{eq:spec_resolution}, the spectral resolution ($\delta\lambda$) depends on the lenslet pitch ($w$), the lenslet focal reduction ($F_{lens}/F_{tel}$),  demagnification through the collimator and camera optics  ($f_{cam}/f_{col}$), anamorphic demagnification from the grating ($r$) \citep{Schweizer1979}, and the angular dispersion ($d\beta/d\lambda$).  

\begin{equation}
\label{eq:spec_resolution}
\delta \lambda =  w^{\prime}r\frac{f_{cam}}{f_{col}}\frac{d\lambda}{dl} = \frac{wr}{f_{col}}\frac{F_{lens}}{F_{tel}}\frac{d\lambda}{d\beta}
\end{equation}

We isolate terms where the diffraction grating contributes to spectral resolution in Equation \ref{eq:spec_resolution}. The product of the logarithmic angular dispersion ($\frac{d\beta}{d\log{\lambda}}$) and the anamorphic factor ($\frac{1}{r}$) is often used as a resolution merit function for the diffraction grating \citep{Bershady2009}.  As mentioned in Section \ref{sec:diffraction}, fine tuning the grating fringe slant directly influences the spectral resolution by inversely changing the anamorphic demagnification and the angular dispersion. A negative (clockwise) fringe slant increases the angle of diffraction at Bragg incidence, improving the angular dispersion. A positive fringe slant (counterclockwise) provides greater anamorphic demagnification at the cost of lowering the angular dispersion. There are instances where we may benefit from a specific fringe orientation, balancing these two factors, and the CH$\alpha$S VPH grating can be installed with the fringe slant in either a positive or negative orientation.

The remaining terms in Equation \ref{eq:spec_resolution} are optimized to improve the spectral resolution while maintaining high S/N. Note that the camera design is slightly faster than the collimator (Table \ref{table:design}), with a shorter focal length ($f_{cam}$) that de-magnifies the field of view and improves the linear dispersion. The focal reduction ($F_{len}/F_{tel}$) provided by the microlens array is set to approximately match the telescope aperture ratio to the collimator aperture ratio. The lenslet pitch ($w$) is selected to carefully balance the angular resolution, spectral resolution, and S/N. Decreasing the pitch of the lenslets in the microlens array decreases the angular extent of each lenslet and reduces the size of the spectral resolution element (similar to reducing the slit width in long-slit spectroscopy). This both improves the spatial resolution and boosts the resolving power. The drawback to decreasing the lenslet pitch is that it also results in a reduced the S/N from each lenslet. We discuss this trade-off more in Section \ref{sec:sensitivity} and Section \ref{sec:discussion}. 
 
\subsection{Sensitivity}
\label{sec:sensitivity}
To demonstrate the sensitivity performance, we calculate the predicted minimum detectable intensity CH$\alpha$S can measure while requiring that we achieve a S/N of 10.  

\begin{align}
\label{eq:signaltonoise}
\rm \frac{S}{N} &= \frac{S}{\sqrt{S + S_{sky} + \sigma_{Dark} + \sigma_{RN}}}\\ 
&= \frac{I_{\lambda} \epsilon  A_{g}  \Omega  t} {\rm \sqrt{S + I_{S}(\lambda) \epsilon  A_{g}  \Omega  t  \Delta \lambda + N_{p} D  t +  (N_{p}/N_{b}^{2}) RN^{2}  N_{f}}}    
\end{align}

The emission signal S in units of photons is the product of the line flux ($I_{\lambda}$), the instrument efficiency ($\epsilon$), the geometric area of the telescope ($A_{g}$), the angular extent on the sky ($\Omega$), and the exposure time (t). Noise contribution terms added in quadrature include shot noise from the emission signal, shot noise from the sky-background emission, dark current from the detector, and read noise from the detector. Our sensitivity analysis (Figure \ref{fig:sensitivitymdf}) includes all of these noise sources.  The design goal is to operate in a sky-background limited regime during dark time, and Equation \ref{eq:signaltonoise} then simplifies to: 

\begin{equation}
\label{eq:signaltonoise_skybkglim}
\rm \frac{S}{N} = \frac{I_{\lambda}\sqrt{\epsilon A_{g} \Omega t}}{\sqrt{I_{s}(\lambda)\Delta \lambda}}
\end{equation}  

In Figure \ref{fig:sensitivitymdf} we model the CH$\alpha$S $10\sigma$ minimum detectable intensity as a function of angular scale. This calculation follows Equation \ref{eq:signaltonoise} assuming an efficiency of $\epsilon = 15\%$. In the sky background limited regime, the signal-to-noise will scale as $\sqrt{\epsilon}$. We show three integration times (1 hour, 10 hours, and 100 hours) and assume that each integration is composed of 900s exposures. Lenslets in each microlens array only cover a few arcseconds in angular size. While it would be exceedingly difficult to reach sensitivities on the order of 1-10 mR in each resolution element, part of the power of using a microlens array is that individual lenslets can be binned together to measure a fixed solid angle on the sky. We measure larger spatial scales by combining the signal from many neighboring lenslets. Combining the signal from multiple lenslets is made easier by the fact that PSF of each lenslet is an image of the pupil and is invariant to the on-sky spatial structure in each lenslet. Although we are probing a coarser spatial scale with this lenslet binning, the measurement maintains a fixed spectral resolution. The S/N improves with increasing scale and integration time in accordance with Equation \ref{eq:signaltonoise_skybkglim}.

\section{On-Sky Performance}
\label{sec:earlycomm}

Using calibration data and on-sky measurements taken during early commissioning in May-June 2021, we have compiled a preliminary performance assessment of CH$\alpha$S on the Hiltner 2.4-m telescope. This is intended as an early validation of the expected performance.

\begin{figure}[ht]
\begin{center}
\includegraphics[width=\columnwidth]{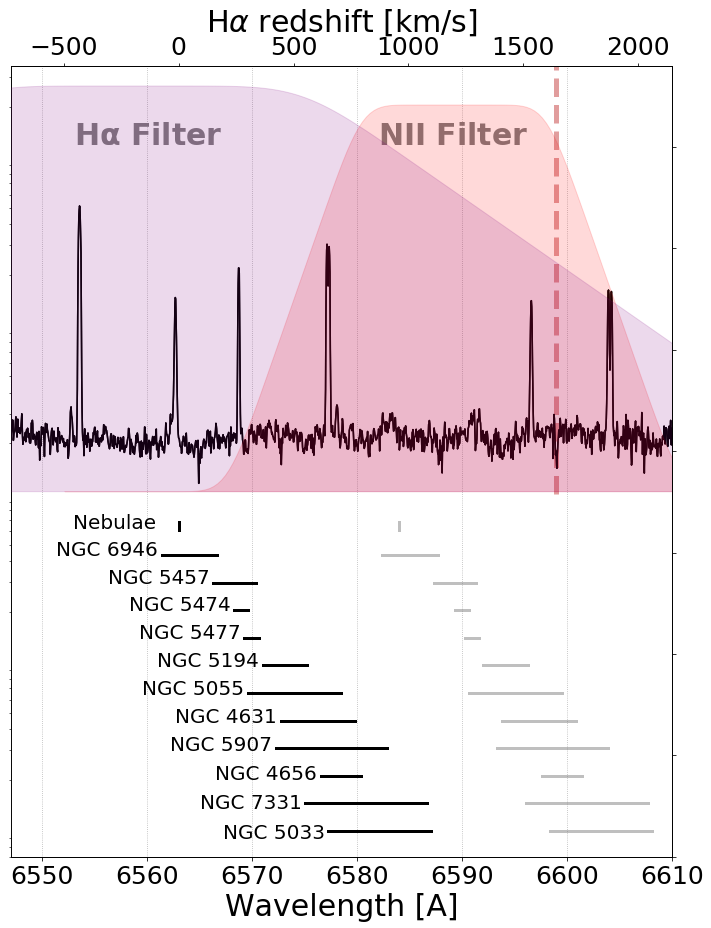}
\caption{CH$\alpha$S early commissioning targets and observational planning. The top spectrum shows the sky background emission lines from the atmosphere. Each black horizontal line is a galaxy centered on its redshifted H$\alpha$ emission. For galaxies in the SING survey, the length of the line corresponds to the H$\alpha$ velocity width  ($W_{20}$) \citep{Kennicutt2003}. For galaxies not in the SINGS survey (NGC 5477, NGC 5457, NGC 5907, and NGC 4656), we use HI velocity width measurements \citep{Davies1980, Rots1980}. Note that the black lines for NGC 5907 and NGC 4656 are full width $\rm \Delta v$ and not $\rm W_{20}$ measurements. The grey lines correspond to the redshifted NII emission (assuming the same width). Nebulae targets are shown as vertical lines at the emitted wavelength of H$\alpha$ and NII. The shaded regions behind the spectrum show the approximate filter response curves for the H$\alpha$ (6 nm FWHM) and NII (3 nm FWHM) filters. The red vertical dashed line marks the emission line wavelength present in the Ne reference lamp. } 
\label{fig:commissioningsummary}
\end{center}
\end{figure}

We summarize the on-sky commissioning targets in Table \ref{table:comtargsummary} and Figure \ref{fig:commissioningsummary}. Here we compare the expected emission from each target/galaxy with the telluric emission from the atmosphere shown in the top panel \citep{Leinert1998}. Each black horizontal line in the bottom panel is a galaxy centered on its redshifted H$\alpha$ emission. The length of the line corresponds to the $W_{20}$ velocity width \citep{Kennicutt2003}. The corresponding grey lines represent the redshifted NII emission (assuming the same width as H$\alpha$). The shaded regions show the approximate filter response curves for the Baader H$\alpha$ (6 nm FWHM) and Astrodon NII (3 nm FWHM) filters. Most of our primary targets have H$\alpha$ emission that is redshifted into the NII filter. %

\begin{figure*}[ht]
\begin{center}
\includegraphics[width=\textwidth]{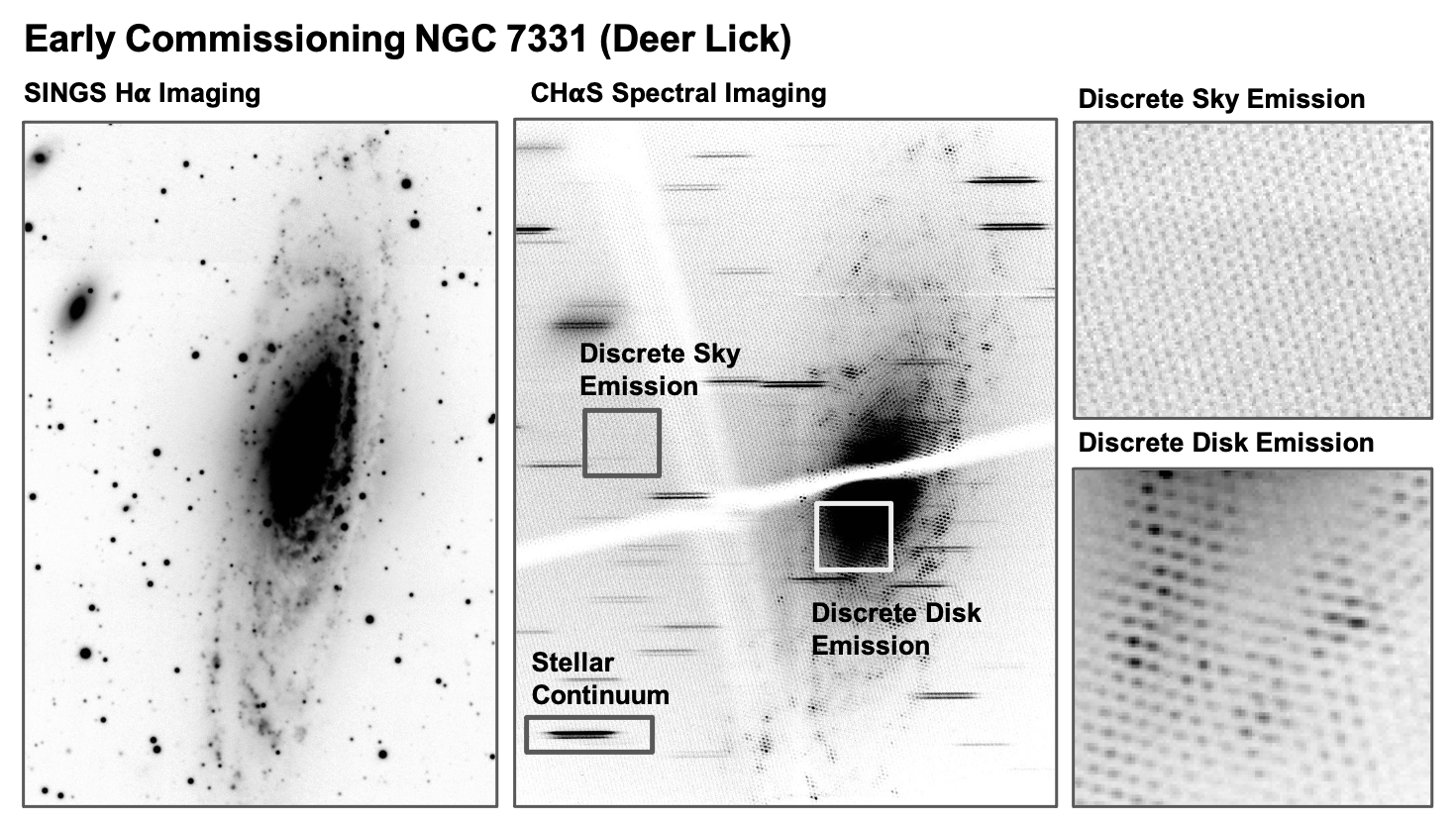}
\caption{Example data from early commissioning. The left panel is an H$\alpha$ image from the SINGS survey for comparison \citep{doi10.26131/irsa424}. The middle panel is a preliminary stack of NGC 7331 (Deer Lick) data taken on 14 June 2021. This is a 1 hour integration composed of 180 second exposures. These observations were taken with a Moon illumination of $15\%$ and a Moon separation of $117^{\circ}$. Sky line emission in the background has not been subtracted, and can be seen in the upper right panel. The cross-hair artifact is a calibration mask installed throughout the commissioning run.} 
\label{fig:commissioningdata}
\end{center}
\end{figure*}

As an example of the commissioning data, we show a stack of the Deer Lick Galaxy NGC 7331 in Figure \ref{fig:commissioningdata}. In many ways this stack resembles a narrowband image of the galaxy, with an H$\alpha$ image from the SINGS survey provided for comparison. On close inspection we see the imprint of the lenslet array in the CH$\alpha$S stack, resembling a pointillism image of the galaxy. This stack is not just an intensity map of NGC 7731, but also a spectral image dispersed along the x-axis. This can be seen in the continuum spectrum dispersed for each star.

The two zoom-in panels in Figure \ref{fig:commissioningdata} show discrete emission. Depending on the gas velocity in a given lenslet bin, each emission line may be Doppler shifted slightly with respect to the systemic velocity. Small wavelength shifts are dispersed through the grating and the pupil image for each lenslet is slightly offset from the lenslet center along the spectral direction. These pixel-scale offsets in the emission line position with respect to the centroid of each lenslet are measured via comparison with a rest-frame image of a calibration lamp. This effect can be seen in the discrete disk emission. Note how the pupil images do not form a perfect grid pattern and instead looks distorted, reflecting velocity offsets in the gas at this location. The discrete sky emission comes from the atmosphere at a nearly uniform velocity and therefore has a constant Doppler shift with respect to the microlens hexagonal grid pattern. Accordingly, we see the sky emission as a uniform image of the microlens array. 

\begin{figure*}[ht]
\begin{center}
\includegraphics[width=\textwidth]{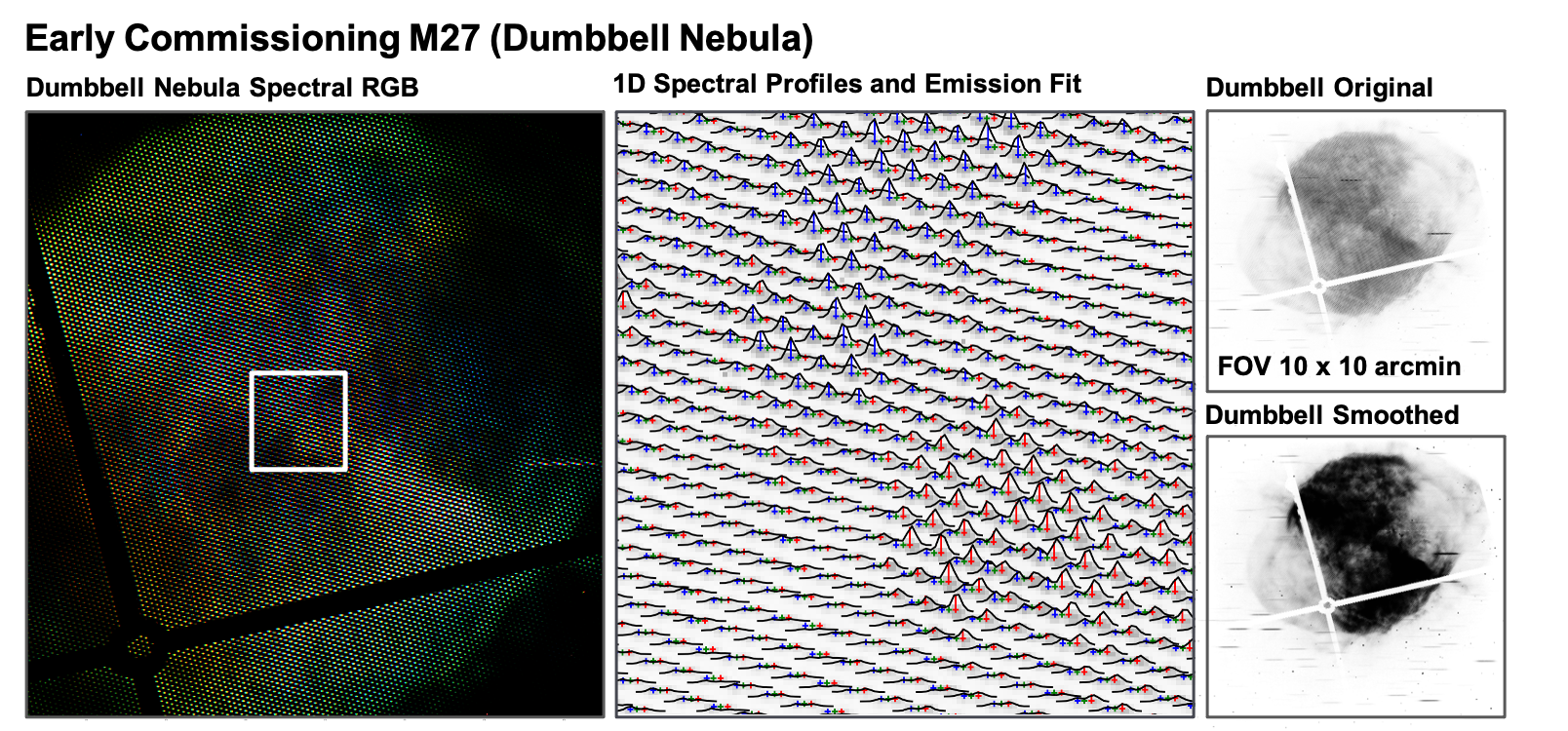}
\caption{Example data from early commissioning. The left panel is an RGB image of the CH$\alpha$S spectra colored by velocity. This map has a resolution of $\pm 2$ pixels or $\pm 60$ km/s. If we zoom in on the center of the nebula and create a plot of the 1D spectral profiles (middle panel) we see the expansion of the nebula. Emission lines in each 1D profile have been fit and colored based on if they are redshifted or blueshifted. In the rightmost panels we see that the Dumbbell nebula nearly fills the $10' \times 10 '$ FOV. The smoothed CH$\alpha$S Data resembles a narrowband H$\alpha$ image of the nebula. The cross-hair artifact is a calibration mask installed throughout the commissioning run.}
\label{fig:commissioningdata2}
\end{center}
\end{figure*}

In addition to the list of galaxy targets in Table \ref{table:comtargsummary}, we have used CH$\alpha$S to observe a variety of nebulae during early commissioning. Wide field spectral imaging with CH$\alpha$S is great for mapping large nebular regions such as planetary nebulae, emission nebulae and supernova remnants. CH$\alpha$S maps these bright targets extremely efficiently, allowing us to tile large regions with only a few exposures per pointing. These targets often fill the full field of view, and are excellent for instrument calibration.  In Figure \ref{fig:commissioningdata2} we show CH$\alpha$S images of the Dumbbell Nebula. Like many nebulae, the Dumbbell Nebula hosts interesting velocity structure, and we highlight the capability of CH$\alpha$S to map this structure in a single 180s exposure with a velocity resolution of $\sim 30$ km $\rm s^{-1}$. A spectrum is extracted for each lenslet using the calibration lamp as a weighted template to find the emission peaks. We produce an RGB image of the CH$\alpha$S spectra colored by velocity (with steps of 30 km $\rm s^{-1}$) to demonstrate the stunning visual properties of this data. We plot 1D profiles for each spectrum at the center showing both single and multi-component emission lines from gas expanding away from the central white dwarf at velocites of $\pm30$ km/s. In future runs, CH$\alpha$S will be capable of collecting similar spectral maps in multiple optical emission lines, probing shock structure and the strength of various ionization mechanisms in a collection of a few short exposures. 

\begin{deluxetable}{llc}
\tablecaption{Primary target integration summary from early commissioning May/June 2021. Each integration is composed of 180 second exposures. We require a Moon illumination of $\leq 50\%$ and a Moon separation of $\geq 55^{\circ}$. }
\label{table:comtargsummary}
\tablehead{
\colhead{Target} & \colhead{Alternate Name} & \colhead{Total Integration}  
}
\startdata
NGC 4631 & Whale Galaxy & 3.0 hrs\\
NGC 4656 & Whale Friend Galaxy & 2.1 hrs\\
NGC 7331 & Deer Lick Galaxy & 4.1 hrs\\
NGC 5906 & Cat Scratch Galaxy & 5.05 hrs\\
NGC $6946^{\dag}$ & Fireworks Galaxy\tablenotemark{$\dag$} & 1.2 hrs\\
NGC 5033 & Dungeness Galaxy & 1.85 hrs\\
\enddata
\tablenotetext{$\dag$}{Observed in H$\alpha$ filter}
\end{deluxetable}

\subsection{Comissioning Data Analysis} 
The raw CH$\alpha$S data is a flattened spectral cube collected in a single exposure without the need to scan in frequency space. In order to build up long integration times we combine a set of exposures from each night into a stacked image. The stacking/image registration process uses the cross correlation to calculate sub-pixel offsets between exposures and align them appropriately. Cosmic rays are removed from each image in the stack by looking for Laplacian/sharp edges following the prescription in \cite{VanDokkum2001}.  We are working on a global solution for the spatio-spectral instrument response across the full field of view. Flat field correction, sky background subtraction, and photometric calibration have not been applied to Figure \ref{fig:commissioningdata}. 

The model for spectral extraction in our science data is based on the location of monochromatic light through each lenslet, which is determined using either a calibration lamp (Figure \ref{fig:commissioningdata_calib}) or a low-velocity nebula. These locations are then shifted, using the dispersion solution, to the expected location of H$\alpha$ at the systemic velocity of the target galaxy, establishing initial centers for the analysis of lenslet spectra on the detector. We extract a spectrum at each of these locations; however, overlap with adjacent lenslets results in a quasi-periodic multi-peaked spectrum. Using a velocity prior, typically an HI map, we weight the peak detection in the extracted spectrum by the expected Doppler shift, limiting it to a small velocity range around the prior velocity.

Cross-talk between neighboring lenslets is partially mitigated by setting an appropriate lenslet array rotation given the hexagonal packing geometry, but there is still spectral overlap to contend with in both the dispersion and cross-dispersion directions.  The amount of overlap is dependent on the selected bandwidth, the lenslet array pitch and cross-dispersion profile and the (optimized) lenslet array rotation (See Figure \ref{fig:microlens}).  As discussed in \citet{Bacon2001}, there is a significant tradeoff to be considered between maximizing signal and increasing random or systematic noise due to overlap.  Since \chas~ is primarily targeting a single emission line, the bandwidth may be optimized for specific targets using ultra-narrowband filters. Additionally, the design of \chas~ allows for masking of the microlens array that can reduce overlap through loss of sky coverage. For the commissioning runs, we have not used masking or ultra-narrowband filters and leave more extensive discussion of the full set of tradeoffs and their impacts on S/N for future work. 

For the commissioning implementation, as seen in Figure \ref{fig:commissioningdata}, narrow/individual emission lines can be identified. We estimate the spectral overlap between lenslets using the non-sequential ray trace model. For narrowband  $3 \rm \AA$ emission (e.g. Doppler broadened spectral lines), we find that $<5\%$ of the flux in a measured spectrum comes from overlap. Still, confusion may arise from continuum emission, multiple velocity components, or overlap with the dispersed sky lines. For continuum emission with a filter bandpass of $20 \rm \AA$, the fractional flux in each spectrum that comes overlap is approximately $66\%$. This overlap fraction is in agreement with measured noise from the sky continuum background shown in Figure \ref{fig:noise}.

\subsection{Optical Performance}
\label{subsec:opticalperformance}
\begin{figure*}[ht]
\begin{center}
\includegraphics[width=\textwidth]{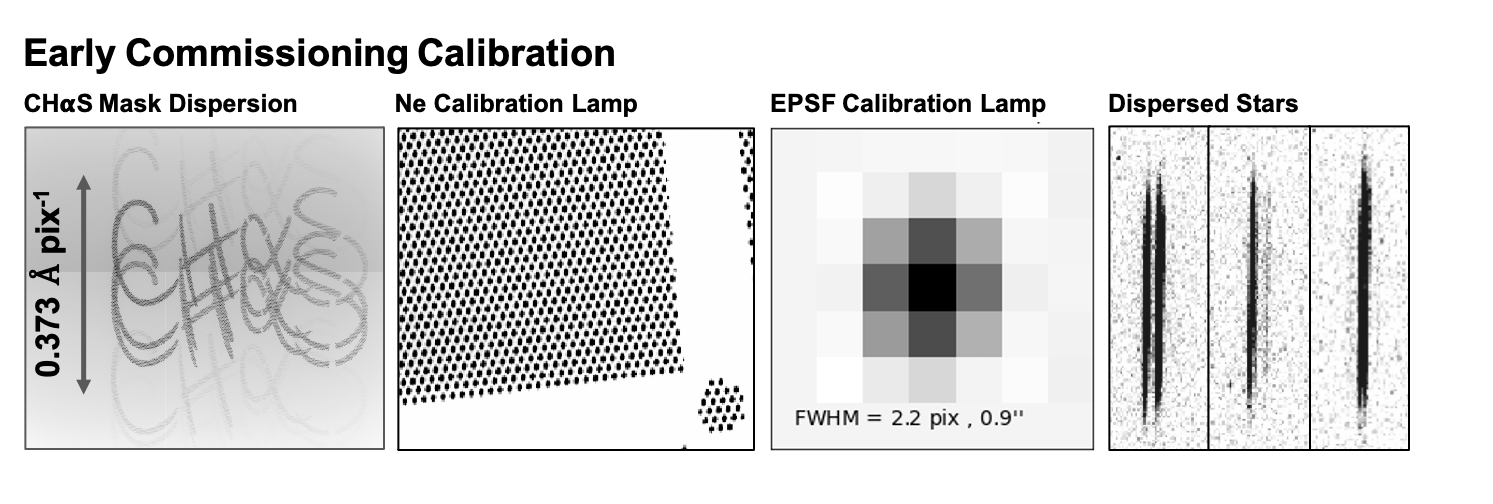}
\caption{(left) Calibration Lamp in H$\alpha$ filter with the CH$\alpha$S mask in front of microlens array. This demonstrates the dispersion along the spectral axis, where each image of the mask is a line in the Ne lamp. (middle left) Ne calibration lamp in the NII filter. This is a zoomed-in single quadrant of the lamp, showing the individual lenslets and the cross-hair mask. (middle right) The effective point-spread function (EPSF) of the calibration lamp lenslet pupil images. (right) Stars dispersed in the NII filter in a single image (at the same telescope focus) hitting different combinations of lenslets. In the broader H$\alpha$ filter, stars dispersed in the CH$\alpha$S field show prominent absorption features. } 
\label{fig:commissioningdata_calib}
\end{center}
\end{figure*}

During commissioning calibration lamps were routinely collected for all filters. The optical commissioning performance can be measured across the field of view using these calibration lamps, an example of which is shown in Figure \ref{fig:commissioningdata_calib}. This Ne lamp is taken in the NII filter and is a direct measure of the spectrograph PSF (without the telescope) in discrete emission at 6598 $\AA$ . We measure an average circular FWHM spot size of 1.8 pixels in the commissioning lamps or 54 microns. This is similar to the effective PSF in Figure \ref{fig:commissioningdata_calib} and is on par with predictions in Section \ref{subsec:opticalperformance}. There is a slight astigmatism in the PSF, most prominent at the edges of the field.  This is most likely due to a tilt in the co-planar alignment of the camera focal plane with the detector CCD \citep{Bacon1995}.

We apply a World Coordinate System solve to the on-sky commissioning images using stellar catalogs in each pointing. This solve aligns the stellar catalog positions with the wavelength of the H$\alpha$ emission line redshifted to the systemic velocity of the galaxy. This provides the best average alignment between CH$\alpha$S emission features and narrowband imaging. In both directions, the telescope plate scale of $\rm 0.34 '' \ pix^{-1}$ is magnified by the ratio of the CH$\alpha$S collimator to camera focal length. We recover a plate scale at the detector of $\rm 0.457 '' \ pix^{-1}$ in the cross-spectral direction and $\rm 0.421 '' \ pix^{-1}$ in the spectral direction. In the spectral direction, the anamorphic magnification from the grating (installed with a negative fringe slant) accounts for the additional factor of 1.09 in the plate scale measurement (See Table \ref{table:gratingparams}). 

In order to recover an estimate for the angular spatial resolution, we measure the average separation between detected lenslets in our commissioning data. Here we use images with a WCS solve and very little expected velocity structure. Our model predicts that the on-sky angular resolution for the ML250 array should be $2.83'' $. We measure an angular resolution of $2.77'' \pm 0.06 ''$, consistent with this expectation. 

\subsection{Spectral Performance}

We measure the linear spectral dispersion using two independent methods. (1) Using the broader (6 nm) H$\alpha$ filter, we detect multiple emission lines from the Ne lamp (Figure \ref{fig:commissioningdata_calib}). The separation between lines at known wavelengths can be used to solve for the dispersion. (2) Bright M-type stars dispersed in the commissioning data show spectral absorption features in the NII filter. These absorption lines can be compared with existing high resolution spectra to solve for the dispersion. Both methods recover an operational linear dispersion of $\rm 0.373 \pm 0.001 \ \AA \ pix^{-1}$. We use this dispersion measurement and the optical performance assessed in Section \ref{subsec:opticalperformance} to estimate the spectral resolution. Given a FWHM of 1.8 pixels and a dispersion of $\rm 0.373 \ \AA \ pix^{-1}$ we achieve a commissioning spectral resolution of $\rm 0.67 \ \AA$ or $\rm 30 \ km \ s^{-1}$. The ML125 array results in spots with a measured FWHM of 1.01 pixels, corresponding to a spectral resolution of $\rm 0.377 \ \AA$ or $\rm 17 \ km \ s^{-1}$.

\subsection{Mechanical Performance} 
\label{sec:commechperformance} 

The CH$\alpha$S mechanical design must maintain minimal flexure over long integration times in order to avoid blurring/degrading the spectral imaging quality. The mechanical performance can be assessed using two critical flexure measurements: (1) flexure in the entrance optics and relay system (2) flexure in the spectrograph including the collimator, camera, and pivot. Entrance flexure blurs the re-imaging of the telescope focal plane on the surface of the microlens array and degrades the spatial resolution. Spectrograph flexure causes the microlens array image to shift around on the detector. It is axis dependent; a drift along the spectral direction degrades the spectral resolution, while a drift in in the cross spectral direction degrades the spatial resolution element. CH$\alpha$S does not have imposed pointing constraints when attached to the telescope, so it experiences a full range of gravity vectors during normal operation. As a commissioning goal, we aim to reduce the acceptable drift due to flexure to $<50\%$ of the spectral imaging  performance. This sets the following requirements using the ML250 configuration: 

\begin{enumerate}
    \item Entrance flexure (in the re-imaged telescope focal plane) should not exceed half a lenslet or $1.3''$.
    \item Spectrograph flexure should not exceed half a spectral resolution element or 30 microns (1 pixel in 2x2 binning). 
\end{enumerate}

\begin{figure}[ht]
\begin{center}
\includegraphics[width=1.05\columnwidth]{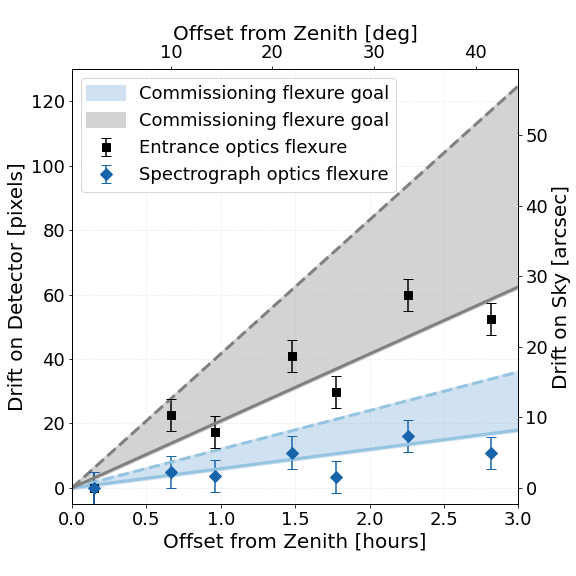}
\caption{Early commissioning flexure measurements. The blue points are the measured entrance flexure and the black points are the measured spectrograph flexure. Each measurement is the magnitude of the drift calculated by summing flexure along the spectral direction and flexure along the cross-spectral direction in quadrature. The shaded regions represent the expected drift due to flexure over a range of exposure times spanning from 300 seconds (dotted line) to 600 seconds (solid line). These are calculated using the commissioning flexure goals outlined in text. }
\label{fig:flexure}
\end{center}
\end{figure}

Early commissioning measurements of the instrument flexure were designed to probe both entrance flexure and spectrograph flexure. Reference cross-hair masks were installed on both the filter and microlens array. The telescope was slewed incrementally in RA (the flexure is much smaller in Dec), and the shift in both cross-hair images was measured at the detector. We summarize the results in Figure \ref{fig:flexure}. Discrete points are the measured entrance (blue) and spectrograph (black) flexure. Each measurement is the magnitude of the drift calculated by summing flexure along the spectral axis and flexure along the cross-spectral axis in quadrature. However, we note that the flexure in the entrance optics is primarily in the spectral direction and the flexure in the spectrograph optics is primarily in the cross-spectral direction. The shaded regions represent the expected drift due to flexure using the commissioning requirements over a range of exposure times spanning from 300 seconds (dotted line) to 600 seconds (solid line). These measurements along the equatorial axes and the flexure measured during on-sky observations show a very similar trend. We find that the current flexure is consistent with our limiting requirements at an exposure time of around 600s. Early commissioning runs used a very conservative exposure time of 180s. Exposures can be lengthened to $\rm 180s < t_{exp} < 600s$ in subsequent runs without degrading the spectral imaging beyond commissioning requirements. These numbers become more stringent by an additional factor of $2$ if we swap out the ML250 microlens array for the smaller pitch, higher resolution ML125 array. Ultimately, we aim to make the flexure requirements for CH$\alpha$S a factor of $5$ tighter than the commissioning requirements. This would reduce the acceptable drift due to flexure to $<10\%$ of the spectral imaging performance.

\subsection{Sensitivity Performance} 
\begin{figure}[ht]
\begin{center}
\includegraphics[width=1.1\columnwidth]{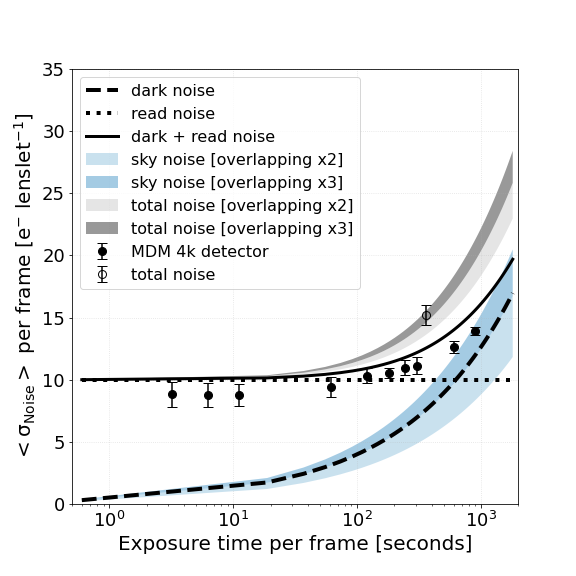}
\caption{CH$\alpha$S noise budget including sky background and MDM 4K detector noise. We plot the standard deviation $\sigma_{N}$ of each noise source per lenslet per frame as a function of the exposure time per frame. We assume a CCD binning of $2\times 2$ and a spot size per lenslet of 60 $\mu$m (corresponding to an area per lenslet of $\sim 4$ pixels). The expected read noise (black dotted line) is constant per frame but remains a function of the number of frames per total integration time. The expected dark noise (dashed black line) increases with the exposure time per frame. These sources are added in quadrature to estimate the total detector noise (solid black line). Discrete points are detector dark data from our 2021 commissioning tests. The MDM 4K detector is a mosaic of 4 CCDs. Each data point is the average detector noise (bias subtracted) at the center of each quadrant, and the error bars are the standard deviation between quadrants. Similarly, the total noise shown as a discrete open circle is measured by taking the lenslet-to-lenslet standard deviation in a blank sky frame. \label{fig:noise}}
\end{center}
\end{figure}

The CH$\alpha$S noise budget including the sky background and MDM 4K detector noise is shown in Figure \ref{fig:noise}. Here we plot the standard deviation $\sigma_{N}$ of each noise source per lenslet per frame as a function of the exposure time. This is independent of the total integration time composed of N frames. We use a CCD binning of $2\times 2$ and a spot size of 60 $\mu$m (corresponding to an area per lenslet of $\sim 4$ pixels). The expected read noise (black dotted line) is constant per frame but remains a function of the number of frames per total integration time. The expected dark noise (dashed black line) increases with the exposure time per frame. These sources are added in quadrature to estimate the total detector noise (solid black line). Discrete points are detector dark data from our 2021 commissioning tests. The MDM 4K detector is a mosaic of 4 CCDs. Each data point is the average detector noise (bias subtracted) at the center of each quadrant, and the error bars are the standard deviation between quadrants. These measurements confirm that the read noise and dark current are similar to the detector specifications and demonstrate the light-tight capabilities of our instrument enclosure in the dome environment. In addition to the detector noise, the sky background noise can have a range of values covered by the blue shaded regions in Figure \ref{fig:noise}. These values assume  a sky background intensity of 2 $\rm R \ \AA^{-1}$ \citep{Leinert1998}. If the microlens array is partially masked and the bandpass is narrow enough that continuum from the sky spectra does not overlap at all, then the sky background is the lower boundary of the light blue shaded region. In this case, CH$\alpha$S is detector-limited by the dark noise in the MDM 4K. However, even with a narrow bandpass of 30 - 60 $\AA$, the continuum from the sky background in neighboring lenslets overlaps by a factor of 2-3 (shown by the darker blue shading). In this nominal configuration without any masked lenslets, the sky background noise becomes comparable to or slightly exceeds the dark noise and CH$\alpha$S is sky-background limited. The total noise (again summed in quadrature) is shown by the shaded gray region over the range of sky background values. We measure the total noise directly (including the sky and detector) in regions of blank sky in the commissioning data. This measurement for a 360s frame is plotted as discrete open circle. It confirms the model used in our signal-to-noise predictions and shows that we are currently operating in a sky-background limited regime.

In addition to the detector noise performance, the sensitivity relies on the full instrument throughput. In Table \ref{table:performance} we outline the expected performance of the telescope, spectrograph and detector combined, predicting a total throughput of $\epsilon = 15\% - 20\%$ without the atmosphere. Early commissioning throughput testing of the spectrograph optics has been completed using a laser at $6563 \AA$ fed into the system (below the filter) with an F/7.5 input to match the speed of the telescope. The intensity of this input was measured separately using a beam splitter to a photometer. We measure a spectrograph efficiency of approximately $15\%$ not including the telescope, filter, or atmosphere. We have also made an on-sky measurement using a bright star placed at the center of the field. We compare with the SDSS r-band magnitude, as this is the closest to our NII filter. This measurement has been corrected for the $75\%$ fill factor of the microlens array as the star is detected over multiple lenslets and about $25\%$ of the flux is blocked by the black chromimum mask between lenslets. We measure an full-system efficiently of $8\%$ in current configuration. This temporary factor of 2 loss in throughput is due a combination of the severely aged telescope coating and the commercial filter performance. We anticipate improvement of these factors and they will not have a long-term affect on the system throughput.

\section{Discussion}
\label{sec:discussion}

\begin{figure*}[ht]
\begin{center}
\includegraphics[width=0.6\textwidth]{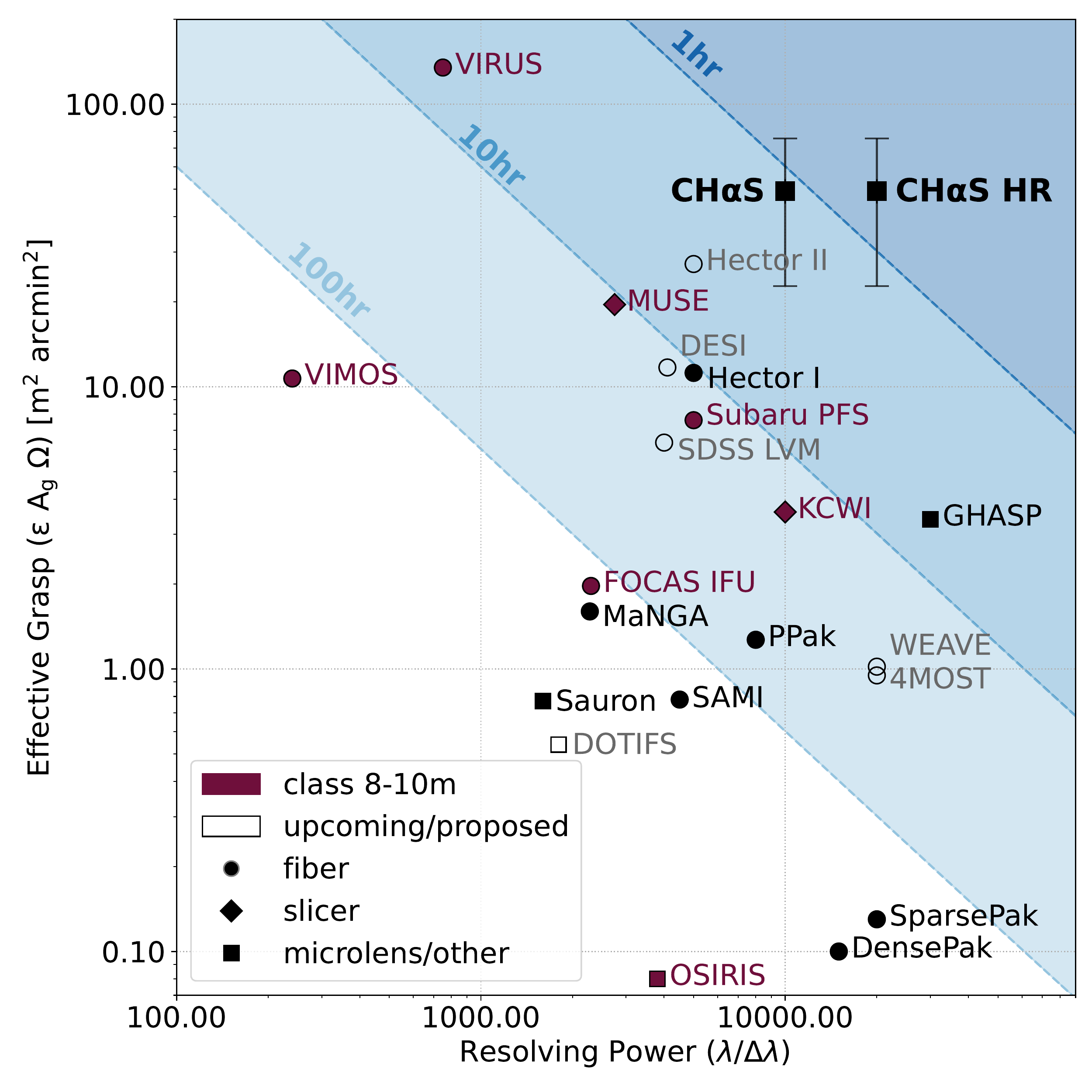}
\caption{A comparison of effective grasp vs. resolving power between integral field spectrographs. Symbols and coloring distinguish the telescope size, instrument commissioning status, and focal plane coupling. When an efficiency measurement is not available we assume $\epsilon = 0.2$. The CH$\alpha$S design (nominal and high resolution) occupies a unique quadrant of this parameter space, maintaining a high grasp and moderate resolving power. Contours are exposure time isochrones showing the grasp/resolution needed for a 10$\sigma$ detection of diffuse H$\alpha$ emission with an intensity of 3 mR over a solid angle equal to the full field of view for each instrument. Shading encompasses instruments that can make this detection within 1 hour (dark blue), 10 hours (medium blue), and 100 hours (light blue). This calculation excludes detector noise for comparison across platforms. It is only valid for diffuse emission, as it assumes the signal is binned over the full field of view.  \label{fig:compare}}
\end{center}
\end{figure*}

We discuss CH$\alpha$S in comparison with other IFU designs and review the trade-off decisions considered while optimizing the instrument for sky-background limited observations of diffuse signal. The key advantages of the CH$\alpha$S design can by summarized by stating that CH$\alpha$S is a large grasp, high throughput instrument with moderate spectral resolution. While there are a variety of different instrument merit criteria, one common parameter space for assessing relative performance between instruments is the combination of coverage vs. resolution \citep{Bershady2009}. In Figure \ref{fig:compare} we plot the effective grasp vs. resolving power, comparing CH$\alpha$S with many excellent existing and planned integral field spectrographs. CH$\alpha$S occupies a unique quadrant of this plot, maintaining both a high effective grasp and a moderate resolving power.

Maintaining a high grasp and moderate spectral resolution are compelling instrument criteria for mapping discrete, ultra-low surface brightness emission. This combination directly drives the scientific capabilities of CH$\alpha$S. We demonstrate this concept in Figure \ref{fig:compare} by overlaying exposure time isochrones representing the grasp/resolution needed for a 10$\sigma$ detection of diffuse H$\alpha$ emission with an intensity of 3 mR over a solid angle equal to the full field of view for each instrument. Shading encompasses instruments that can make this detection within 1 hour (dark blue), 10 hours (medium blue), and 100 hours (light blue). The dark blue contour segment requiring the shortest exposure time resides in the top right corner of the plot, favoring high grasp/high resolution configurations. The CH$\alpha$S effective grasp is driven by the instrument's large field of view ($10' \times 10')$.  While many integral field spectrographs are paired with $4-8$ meter class telescopes, prioritizing high spatial resolution over wide-field imaging, CH$\alpha$S is paired with a 2.4-meter telescope operating over a wide field of view.
Maintaining a moderate spectral resolution allows for the detailed sky background rejection essential for faint sky-background limited observations; sky background noise at a specific wavelength is reduced by attaining a smaller spectral resolution element (Equation \ref{eq:signaltonoise_skybkglim}). This resolution-driven sensitivity gain is only realized for discrete emission line mapping, where the signal is independent of the bandpass. 
In selecting the spectral resolution, the CH$\alpha$S design carefully balances the spectral vs. spatial information. CH$\alpha$S fits an exceptional number of total ($\rm spatial \ N_{\Omega} \times \ spectral \ N_{R}$) resolution elements on its detector, optimizing the instrument's informational collecting power \citep{Bershady2009}. The trade-off between high grasp and high spectral resolution results in a scale-dependent sensitivity; at the same grasp, an instrument with higher spectral resolution will have greater sensitivity on smaller scales. This occurs because the S/N in the sky-background limited regime relies strongly on the grasp per spectral resolution element and not solely on achieving unparalleled grasp. We leave for future work a comparison with alternative spectrograph designs such as Fabry-P\'{e}rots or Interferometers, some of which may achieve a similarly high grasp and moderate spectral resolution. Because of the complexities of these instruments, there are trade-off assessments and questions regarding their backgrounds that need to be included in a more detailed comparison.

\begin{figure*}[ht]
\begin{center}
\includegraphics[width=0.75\textwidth]{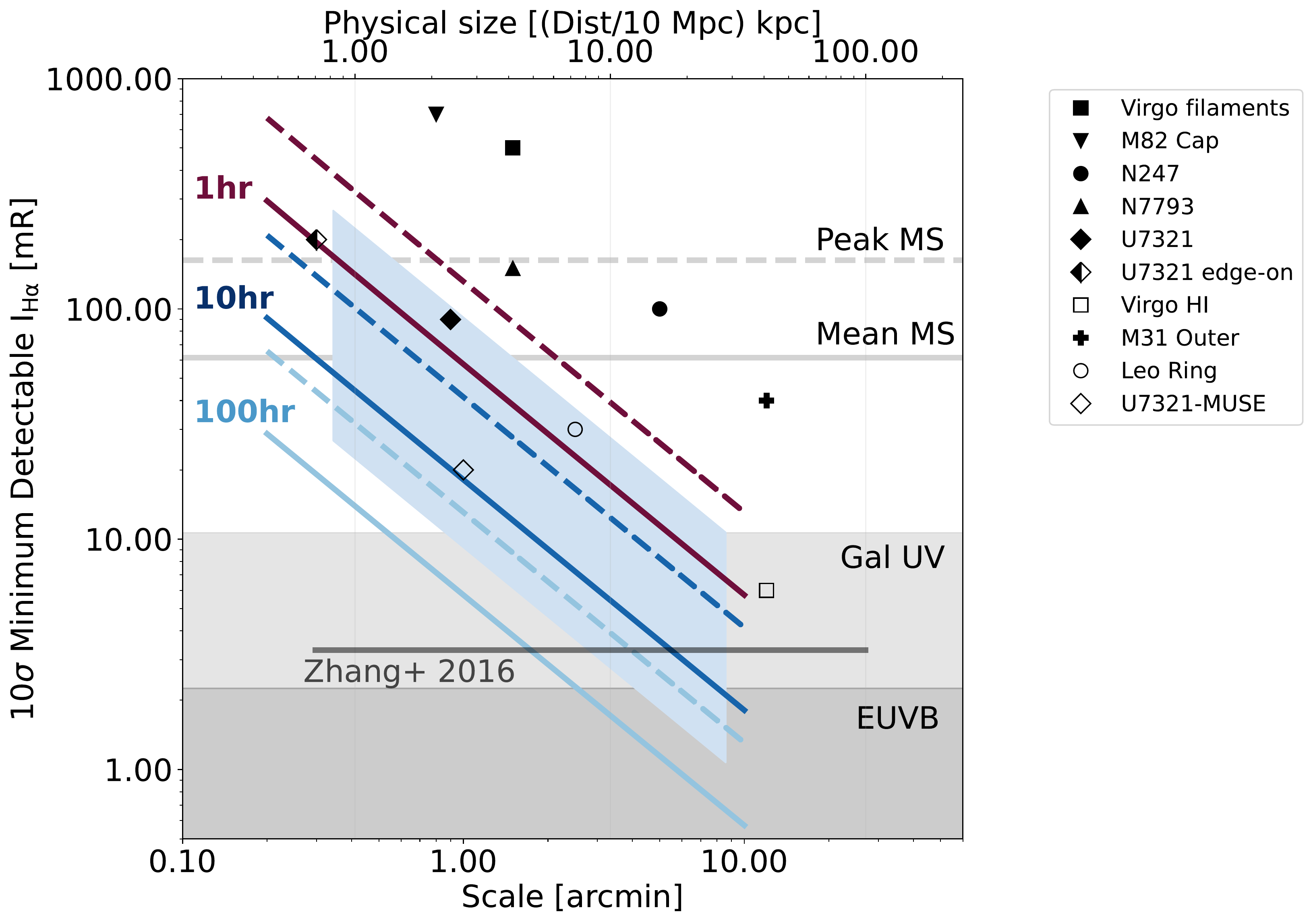}
\caption{The predicted minimum detectable H$\alpha$ intensity (S/N = 10) as a function of angular scale at three different integration times: 1 hr, 10 hr, 100 hr. Soid lines are calcuated with an efficiency $\eta = 15\%$ and dashed lines are calculated with an efficiency $\eta = 6\%$. We compare these sensitivity limits with the expected emission flux from the ultra-faint extragalactic UV background (EUVB) and the galactic UV background. We show that CH$\alpha$S is able to reach these sensitivity limits within 10-100 hours of integration. }
\label{fig:sensitivitymdf}
\end{center}
\end{figure*}

\begin{equation}
    \label{eq:EM}
    \rm EM = \frac{\Gamma_{z=0} N_{HI}}{\alpha_{B}(T)} = 2.75 I_{H\alpha} T_{4}^{0.9} \ cm^{-6} \ pc
\end{equation}

To demonstrate the scientific capabilities of CH$\alpha$S we compare the CH$\alpha$S predicted sensitivity limits with emission observed in the local universe (Figure \ref{fig:sensitivitymdf}).  Observational measurements and upper limits are shown as discrete symbols. Bright H$\alpha$ features include the M82 Cap \citep{Lehnert1999}, the Virgo filaments \citep{Kenney2008}, edge-on galaxies \citep{Christlein2010}, NGC 7793 \citep{Dicaire2008}, UGC 7321 \citep{Adams2011}, NGC 247 \citep{Hlavacek-Larrondo2011}, the outer disk of M31 \citep{Madesen2001}, the Leo Ring \citep{Donahue1995}, and the Virgo HI cloud \citep{Weymann2001}. The dashed and solid light grey lines around 100 mR represent H$\alpha$ emission observations and surface brightness predictions for the Magellanic stream/bridge \citep{Bland-Hawthorn2007, Barger2013}. CH$\alpha$S is capable of making most of these detections on arcminute scales in $\lesssim 1$ hour, and observations of many of these targets have already been carried out with CH$\alpha$S. Emission from the circumgalactic medium is much more tenuous, requiring measurements that push the boundaries of low surface brightness imaging. Faint emission from ionized hydrogen in the outskirts of galaxies and well into the CGM is a probe of the local ionizing flux \citep{Sunyaev1969}. The predominant sources of ionizing radiation in the CGM is the UV continuum from the galactic disk and the photo-ionizing extragalactic UV-background (EUVB). Constraints on the low-redshift (z = 0) EUVB photoionization rate range from $\Gamma \lesssim 2-8 \times 10^{-14} \rm \ s^{-1}$ \citep{Adams2011, Madau2015,  Fumagalli2017, Caruso2019}. The H$\alpha$ emission intensity (quoted in Rayleigh) scales as the emission measure given in Equation \ref{eq:EM}. We use this approximation to calculate a rough estimate of the CGM surface brightness assuming a warm gas temperature $\rm T = 10^{4} \ K$, an EUVB ionization rate $\Gamma = 4 \times 10^{-14} \rm \ s^{-1}$, and a column density $\rm N_{HI} \geq 10^{17} \ cm^{-2}$ for case B recombination ($\alpha_{B}$) in optically thick Lymann Limit systems. With these assumptions, warm ionized gas far into the CGM should produce an exceedingly faint emission signal on the order of a few mR. The \cite{Zhang2016} statistical detection of circumgalactic H$\alpha$ emission achieved by stacking millions of SDSS sightlines has a similar flux of 3 mR.  Higher column density gas or gas ionized by stronger UV continuum flux from the galactic disk would produce brighter emission on the order of 10 mR. For example, high velocity clouds in the halo of the Milky Way ionized by the Galactic disk exhibit H$\alpha$ emission intensities of 30-70 mR \citep{Putman2003}. Following a calculation similar to Equation \ref{eq:EM} and assuming a 50 kpc region, the light blue shaded parallelogram region shows the predicted CGM emission intensity for a range of hydrogen densities $\rm log \  n_{H}/cm^{-3} = -4.5 \ to \ -3.5$ corresponding to an overdensity of  $\delta \sim 100 - 1000$ with clumping on scales of $2-25$ kpc. We show that CH$\alpha$S can probe the majority of this phase space within a few tens of hours.  Within 10-100 hours, CH$\alpha$S is capable of reaching a sensitivity faint enough to place our own constraints on the EUVB photoionization rate. A 50-100h exposure with CH$\alpha$S would be the deepest H$\alpha$ image and velocity field ever obtained, reaching a surface brightness of a few mR on scales of a few arcmin. A CH$\alpha$S science campaign is currently underway, building up 10-50 hour observations on a small sample of Fall/Spring galaxy targets.

Achieving accurate sky subtraction will require a precise and stable flat-field and careful control of systematic effects.  With respect to flat-field, since we are combining many hours of images (corresponding to $>$100 exposures), and combining spectra within a field (e.g. $>$100 spectra per ‘sky bin’), random and spatially uncorrelated flat-field uncertainties will decrease as sqrt(N).  Because of this, flat-field accuracies of 1-10\% should be sufficient and feasible.  Other aspects will require careful calibration and analysis, including bias subtraction, removal of continuum from background point sources, stray and scattered light, detector ghost images, correction for distortion, flexure and sky and spectral variability.  These may reduce the accuracy of the sky subtraction and the reliability of the measured signal and will be addressed in future work that seeks to reach the faintest limits.

\section{Summary}
\label{sec:summary}
\begin{enumerate}
    \item CH$\alpha$S is a narrowband (3 nm), moderate resolution (R $\sim 10,000$), wide-field ($10^{\prime} \times 10^{\prime}$) integral field spectrograph designed to detect faint optical emission from diffuse gas in the local universe (z $< 0.01$). 
    
    \item CH$\alpha$S has a catadioptric design optimized to perform over a 400 - 900 nm operational wavelength range. It is capable of targeting a broad number of optical emission lines including H$\alpha$, H$\beta$, [SII], [NII], [O III], [O II], and [O I]. 
    
    \item  CH$\alpha$S collects a total of $1.75\times 10^{6}$ resolution elements ($\rm N_{\Omega} \times N_{R}$) in a single exposure while maintaining a high grasp and moderate resolving power. This an ideal combination for efficiently mapping discrete, ultra-low surface brightness emission. 
    
    \item A 50-100 hour integration with CH$\alpha$S would be the deepest H$\alpha$ image and velocity field ever obtained, probing emission with a surface brightness of a few milli-Rayleigh while maintaining a S/N of 10.
    
    \item CH$\alpha$S is deployed on the Hiltner 2.4-m telescope at MDM Observatory. We have completed a successful early commissioning at the observatory, and we are working towards making CH$\alpha$S a facility instrument available to the MDM consortium. The instrument remains a testbed for new design concepts and a teaching opportunity for many student-led upgrades.

\end{enumerate}
The Circumgalactic H$\alpha$ Spectrograph is funded by NSF AST-1407652 (ATI). Graduate student Nicole Melso was supported with an NSF GRFP under grant number DGE-1644869, and graduate student B\'{a}rbara Cruvinel Santiago is supported by NASA FINESST 80NSSC19K1419. CH$\alpha$S is commissioned at the MDM Observatory, operated by Dartmouth College, Columbia University, Ohio State University, Ohio University, and the University of Michigan. We especially thank the MDM Observatory staff Eric Galayda and Tony Negrete for their help in commissioning CH$\alpha$S on the Hiltner 2.4-meter as well as John Thorstensen and Jules Halpern for supportive observatory scheduling. This work is built on the success of protoCH$\alpha$S, a prototype constructed and tested by Erika Hamden and Sam Gordon. We thank Jos\'{e} Manuel Zorrilla Matilla, Julia Blue Bird, Derek Wacks, and Katherine Gonglewski for their help at various stages of this project. The construction of this instrument was done in collaboration with many industry/manufacturing companies listed throughout this paper. Special thanks go to Dominic Speer, Elroy Pearson, Misty Johnson, and Wasatch Photonics for the technical discussions and efficiency modeling that led to the current VPH grating design. Many thanks to Vladimir Leleko at Advanced Microoptics Systems (a$\mu$s) for advising us on the design of custom microlens arrays for this application.  The CH$\alpha$S optical design relies on an NDA with Celestron LLC. We are grateful for their collaboration with with academic institutions and for the support provided by their engineering and sales teams. 
\facility{MDM:2.4m}








\begin{deluxetable*}{llllllllll}
\tablecaption{CH$\alpha$S Offner relay design \label{table:design}}
\tablewidth{700pt}
\tabletypesize{\scriptsize}
\tablehead{
\colhead{} & \colhead{Surface} & 
\colhead{Radius} & \colhead{Thickness} & \colhead{Semi-Diameter} & \colhead{Material} & \colhead{Decenter X} & \colhead{Decenter Y} & \colhead{Tilt X} & \colhead{Tilt Y} } 
\startdata
1 & Object & Infinity & 654.627 & 33.0 & & \\
2 & Coordinate Break & & & & & 0.0 & 0.0 & 8.364 & 0.0\\
3 & Concave Sphere & -609.6 &  & 75.0 & MIRROR & \\
4 & Coordinate Break & & -305.927 & & & 0.0 & 0.0 & 8.364 & 0.0\\
5 & Coordinate Break & & & & & 0.0 & 0.0 & -16.728 & 0.0\\
6 & Coordinate Break & & & & & 0.0 & -3.0 & 0.0 & 0.0\\
7 & Convex Sphere & -310.661 & & 25.4 & MIRROR & \\
8 & Coordinate Break & &  & & & 0.0 & 3.0 & 0.0 & 0.0\\
9 & Coordinate Break & & 305.927 & & & 0.0 & 0.0 & -16.728 & 0.0\\
10 & Coordinate Break & &  & & & 0.0 & 0.0 & 8.5 & 0.0\\
11 & Concave Sphere & -609.6 & & 75.0 & MIRROR & \\
12 & Coordinate Break & & -551.031 & & & 0.0 & 0.0 & 8.5 & 0.0\\
\enddata
\end{deluxetable*}

\begin{deluxetable*}{llll}
\tablecaption{CH$\alpha$S optical design parameters \label{table:design}}
\tablewidth{700pt}
\tabletypesize{\scriptsize}
\tablehead{
\colhead{Parameter} & \colhead{Value} & 
\colhead{Unit} & \colhead{Comments}} 
\startdata
\textbf{Design Requirements} & & & \\
Operational wavelength range & 400 - 900 & nm & broad range of optical emission lines \\
Operational Bandpass & 3 & nm & avoid overlapping spectra \\
\hline
\textbf{Optical Overview} &  & &  \\
Telescope aperture ratio & $f/7.5$ & & MDM Observatory Hiltner 2.4m \\
Offner relay aperture ratio & $f/7.5$ & & \\
Microlens array aperture ratio & $f/2.5$ & & \\
Collimator aperture ratio & $f/2.2$ & & $14"$ Rowe-Ackerman Schmidt Astrograph (RASA 14)\\
Camera aperture ratio & $f/2.1$ & & \\
Collimator focal length & 790 & mm & \\
Camera focal length & 590 & mm & \\
\hline
\textbf{Microlens Array} &  & &  \\
Array dimensions & $68 \times 68 $ & mm & \\
Clear aperture & $66 \times 66 $ & mm & \\ 
Material & Fused Silica & & refractive index n = 1.4563 @ 0.6583 nm \\
Thickness & 2.0 & mm & \\
Aperture ratio & $f/2.5$ & & w/ black chromium mask\\
Lenslet pitch [ML250, ML125] & [0.25, 0.125] & mm & \\
Lenslet clear aperture [nominal, high-res] & $[0.225, 0.113] \pm 0.001$ & mm & \\
Lenslet radius of curvature [ML250, ML125] & $[0.257, 0.129] \pm 3\% $ & mm $^{-1}$ & \\
Effective focal length [ML250, ML125] & [0.563, 0.283] &  mm & thick lens approx. at $\lambda = 0.6583$ nm \\
\hline 
\textbf{VPH Grating} &  & &  \\
Fringe frequency ($\nu$) & 1200 & lines/mm & \\
Fringe slant/Blaze ($\phi$) & 4.5 & deg & positive or negative \\
Dimensions & $[340 \times 290 \times 20]$ & mm & fused silica substrate \\
Bulk index (n$_{2}$) & 1.35 & & \\
Index modulation amplitude ($\Delta n$) & 0.11 & & \\
Effective thickness (d) & 3 & $\mu$m & \\ 
\enddata
\end{deluxetable*}

\begin{deluxetable*}{llllll}
\tablecaption{CH$\alpha$S performance and sensitivity calculations \label{table:performance}}
\tablewidth{700pt}
\tabletypesize{\scriptsize}
\tablehead{
\colhead{Parameter} & 
\multicolumn{2}{l}{Value} & 
\colhead{Unit} & 
\colhead{Comments}\\
& Design & CBE \tablenotemark{$^{\dagger}$} & & 
} 
\startdata
\textbf{Telescope Performance} &  & & &  \\
Primary clear aperture diameter &  2.32 & & m & MDM Observatory Hilter 2.4-m \\
Secondary obscuration diameter & 0.745 & & m  & \\
Geometric area & 37914 & & cm$^{2}$ & \\ 
Plate scale & 11.33  & & $\rm ''/mm$ & Focal Plane of 2.4-m\\
\hline 
\textbf{Imaging Performance} &  & &  & \\
Field of view & 10 x 10 & &  arcmin & \\
RMS spot radius & $< 10$ &  & $\mu$m & Monochromatic point source at spectrograph input \\
Pupil image diameter & [60 (4), 30 (2)] & [57(4), 30 (2)] & $\mu$m (pixels) & [ML250, ML125]\\
Spatial Resolution &[$2.83$, $1.42$] & [$2.77$, $1.42$] & arcsec & [ML250, ML125]\\
Image Plate Scale [x, y] & [0.418, 0.455]  &[0.421, 0.457] & $\rm arcsec \ pix^{-1}$ & Commissioning Measurement \\
\hline 
\textbf{Spectroscopic Performance} &  & & & \\
Spectral resolution $\Delta \lambda$ & [0.69, 0.34] & [0.67, 0.377] & \AA & $\lambda 656.3$ nm [nominal, high-resolution] \\
Velocity resolution & 30 (15) & 30(17) & km $\rm s^{-1}$ & [ML250, ML125] \\
Dispersion & 0.4 & 0.373 & $\rm \AA / pix$ &  \\
Number spectral elements (N$_{R}$) & 35 &  &   & per bandwidth, overlapping \\
Number spatial elements (N$_{\Omega}$) & $[5 \times 10^{4}, 2.3 \times 10^{5}]$ & & & [ML250, ML125]\\ 
\hline 
\textbf{Detector Performance} &  &  & & MDM Observatory B4K Detector  \\
Detector pixel size [angular] & 15 [0.17]& & $\mu$m [arcsec] & unbinned\\
Detector read noise & 5 & 4.4 & $\rm e^{-}$ & \\
Detector dark current & 0.01 & 0.007 & $\rm e^{-} \ pix^{-1} \ sec^{-1}$ & At a temperature of -124 $^{\circ}$C\\
Detector binning &  2 $\times$ 2 &  & pixels & \\
\hline 
\textbf{Predicted Throughput/Efficiency} &  & &  \\
Telescope & 0.77 & & & R = 0.88 per surface, aged\\
Filter & 0.97 & & & 3nm Astrodon NII Filter \\
Relay & 0.86 & & & $R>97 \% \times 5$ optics\\
Mircolens & 0.95 & & & R $\approx 4\%$ convex fused-silica, R $\leq 1\%$ planar AR-coated\\
Filling Factor & 0.75 & & & \\ 
Collimator/Camera & 0.9/0.9 & & &  \\
Grating & 0.9 & & & H$\alpha$ Figure \ref{fig:gratingeff}\\
Detector QE & 0.85 & & & MDM B4K \\
Spectrograph Vignetting & 0.50 & & & \\
System throughput & 0.19 & 0.08 - 0.15 &  & predicted without atmosphere \\
\hline 
\textbf{H$\alpha$ Sensitivity Calculation } &  & &  \\
Sky background intensity & 2 ($1.59 \times 10^{5}$) & & R $\rm \AA^{-1}$ (ph s$^{-1}$ cm$^{2}$ sr$^{-1}$, $\rm \AA^{-1}$) &  \cite{Leinert1998} \\
Exposure time per frame & 625 & & seconds  & stacked to reach full integration time\\
Sky background noise & 50 &  & e$^{-}$ per lenslet per frame &  nominal 250 $\mu$m lenslets w/o overlap\\
Dark current noise & 100 & &  e$^{-}$ per lenslet per frame & nominal 250 $\mu$m lenslets\\
Read noise & 100 & & e$^{-}$ per lenslet per frame &  nominal 250 $\mu$m lenslets, 2 pixel binning\\
\\
\textbf{10$\sigma$ Detection Limit} &  & & See Figure \ref{fig:sensitivitymdf} \\
1 hour integration &  109 (8674) & & mR (ph s$^{-1}$ cm$^{-2}$ sr$^{-1}$) & 30 arcsec angular scale, 900s exposures \\
10 hour integration & 34 (2706) & & mR (ph s$^{-1}$ cm$^{-2}$ sr$^{-1}$) & 30 arcsec angular scale, 900s exposures  \\
100 hour integration & 11 (875) & & mR (ph s$^{-1}$ cm$^{-2}$ sr$^{-1}$)& 30 arcsec angular scale, 900s exposures  \\
\enddata
\tablenotetext{\dagger}{The current best estimate (CBE) is the same as the design goal unless independently measured or modeled}
\end{deluxetable*}

\bibliography{chas}{}
\bibliographystyle{aasjournal}



\end{document}